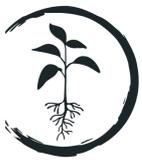
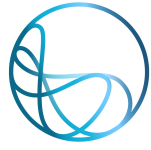
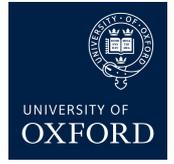
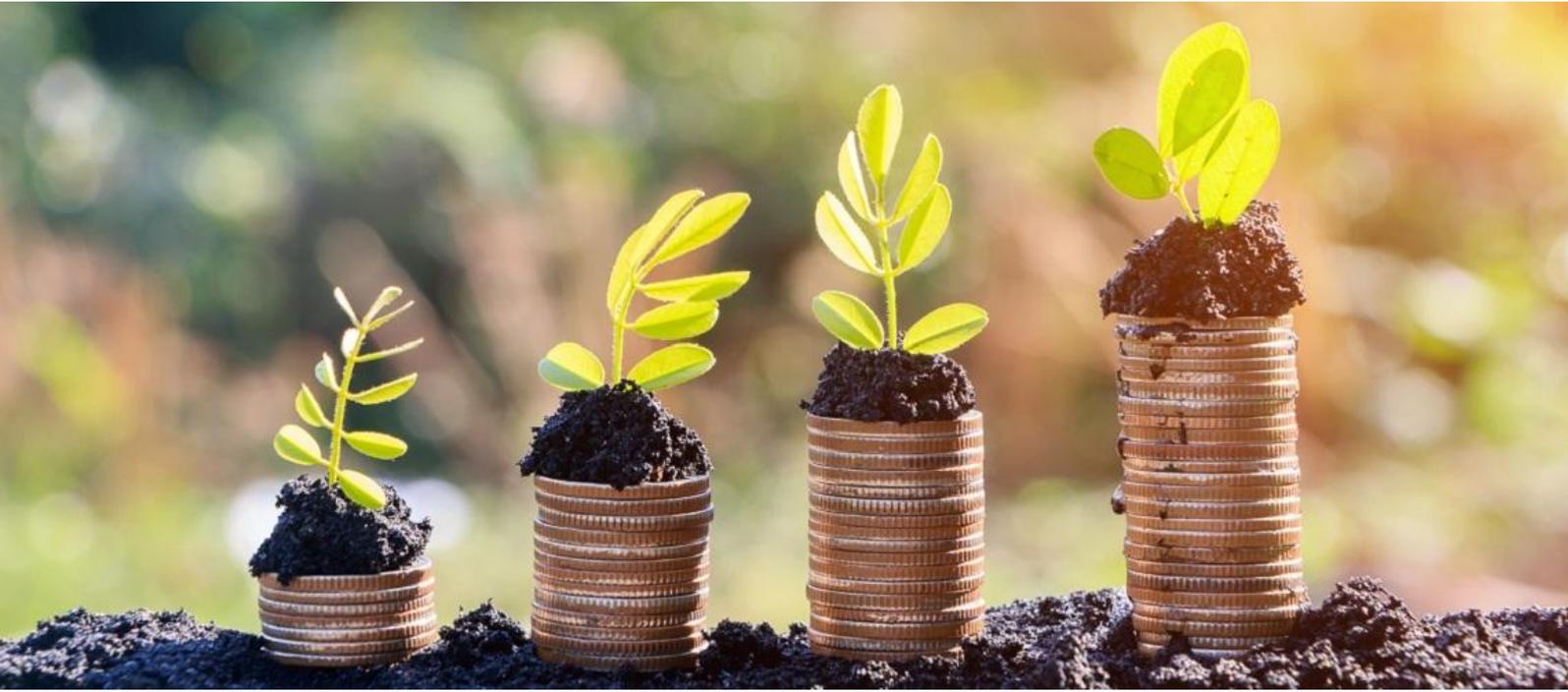

# Assessing the Potential of AI for Spatially Sensitive Nature-Related Financial Risks

Version v1.3

Lead Author: Steven Reece

**Oxford Sustainable Finance Group, Smith School of Enterprise and the Environment, University of Oxford**

**Leverhulme Centre for Nature Recovery
School of Geography
University of Oxford**

# Contributions

This report presents the findings from the *Robust ESG Data for Biodiversity (REDB): Towards a Spatially Sensitive Approach to Sustainable Finance* project that was funded by the Natural Environment Research Council (NERC) as part of the *Integrating Finance and Biodiversity for a Nature Positive Future* call (1st January 2023 to 31st March 2024). The project was led by the University of Oxford, with partners at the University of York, UN Environment Programme World Conservation Monitoring Centre (UNEP-WCMC) and UK Centre for Ecology & Hydrology (UKCEH).

Below we outline the contributions from the respective partners.

### University of Oxford: Steven Reece, Emma O'Donnell

Oxford University led the project, undertook interviews with financial institutions and data providers and subsequently created workflows for the two case-studies. Oxford developed probabilistic value chain models and proposed AI methodologies for both data collection and system level analysis.

### University of York: Felicia Liu

University of York contributed to engaging with various financial stakeholders to scope out existing sustainable investing practices and understanding their diverse data needs to facilitate more effective nature-aligned investing.

### UN Environment Programme World Conservation Monitoring Centre (UNEP-WCMC): Joanna Wolstenholme, Frida Arriaga, Giacomo Ascenzi

UNEP-WCMC contributed to the Brazil use case. They conducted interviews with stakeholders at financial institutions and contributed expertise on nature-related risks and impacts in order to help articulate the AI needs for this case study.

### UK Centre for Ecology & Hydrology: Richard Pywell

UKCEH contributed to the UK Water Utility use case. UKCEH undertook case studies with stakeholders using a range of operational environmental decision support tools and data. This enabled the project team to formulate innovative workflows that incorporated advanced AI algorithms to improve finance and environmental risk assessment and decision making.

Please email any errors in the report to Steven Reece (steven.reece@ouce.ox.ac.uk).

# Executive Summary

There is growing recognition among financial institutions, financial regulators and policy makers of the importance of addressing nature-related risks and opportunities. Evaluating and assessing nature-related risks for financial institutions is challenging due to the large volume of heterogeneous data available on nature and the complexity of investment value chains and the various components' relationship to nature. The dual problem of scaling data analytics and analysing complex systems can be addressed using Artificial Intelligence (AI). We address issues such as plugging existing data gaps with discovered data, data estimation under uncertainty, time series analysis and (near) real-time updates.

This report presents potential AI solutions for models of two distinct use cases: the Brazil Beef Supply Use Case and the Water Utility Use Case. Our two use cases cover a broad perspective within sustainable finance. The Brazilian cattle farming use case is an example of greening finance - integrating nature-related considerations into mainstream financial decision-making to transition investments away from sectors with poor historical track records and unsustainable operations. The deployment of nature-based solutions in the UK water utility use case is an example of financing green - driving investment to nature-positive outcomes. The two use cases also cover different sectors, geographies, financial assets and AI modelling techniques, providing an overview on how AI could be applied to different challenges relating to nature's integration into finance.

This report is primarily aimed at financial institutions but is also of interest to ESG data providers, TNFD, systems modellers, and, of course, AI practitioners.

*Use Case 1: Brazilian Cattle Ranching Sector*

In the Brazil Beef Use case, the focus is on understanding how financial institutions consider biodiversity and nature risks in agricultural sector investments. Deforestation emerged as a significant concern, with increasing litigation risks identified. Our first model adopts a Bayesian approach to analyse relationships within a complex system involving Brazilian banks, meat processors, abattoirs, and farms. The model is used to assess risks related to environmental impact, regulatory compliance, and reputational damage for investors in the supply chain as well as risks to financial return. The report outlines a comprehensive approach to assessing the financial. environmental and reputational risks associated with investments in the Brazilian meat supply chain, leveraging Bayesian modelling and AI-driven data analysis.

*Use Case 2: UK Water Utility Sector*

For the UK Water Utility Use case, discussions revolved around storm overflow damage, water quality improvement costs, and challenges in identifying buffer zones and field margins. The asset manager emphasised the need for innovative approaches, especially in addressing operational constraints and regulatory requirements. Our second proposed model aims to address an optimization problem that balances the expenses related to fines and operational costs against payments for ecosystem services (PES). This approach incorporates a time series framework to allow for strategic adjustments at specific intervals, aligning with bond (and loan) refinancing schedules. Overall, the model integrates financial considerations with environmental goals to optimise investment in nature-based solutions while addressing regulatory requirements and mitigating risks related to water pollution and ecosystem health.

*Key Findings*

Through developing the models ourselves, we were able to engage with all partners in a coherent way on the illustrated problems. We found that the model anchored the conversation and that we were able to apply each organisation's expertise in a meaningful and specific manner without requiring much cross-disciplinary debate. It was particularly challenging yet important to leverage each type of required expertise, be it ecological or financial, as we were unable to identify experts with a deep interdisciplinary understanding of nature-finance relationships.

*Conclusions*

In this report, we explore AI solutions to plug existing data gaps with discovered data, data estimation under uncertainty, time series analysis and (near) real-time updates. Through discussions with stakeholders, we understand how financial institutions consider biodiversity and nature risks in agricultural sector and water utility investments. We describe methods to address an optimization problem that balances the expenses related to fines and operational costs against payments for ecosystem services (PES), representing a "flavour" of potential AI applications in this space.

Without investor validated models, it is difficult to determine exactly how AI would be useful in each use case. With that, our models should be taken with a "pinch of salt" and represent a example of potential AI applications in this space. The relationship between ecological and financial systems, as well as the specific information requirements of financial institutions, must be better understood and articulated for AI to be an appropriate solution. However, we were able to describe potential AI solutions for our two use cases, which should be explored and expanded once models of the relationship between finance and nature become more developed.

The challenge of identifying projects relating to biodiversity and finance that are sufficiently well defined to exploit AI should not be underestimated. In the absence of such projects, we demonstrate where state-of-the-art AI can have impact in both filling data gaps and attributing impact of finance on nature hypothetically. Due to the interdisciplinarity of the topic, a viable use case requires people in finance, supply chains and ecology all interested and willing to engage. We believe that this will happen in the near future. However, there is currently little risk incentive for banks to commit sufficient resources to developing nature-related risk understanding and integration to the level required for risk management or AI implementation.

We have included recommendations for TNFD, systems modellers, financial institutions and ESG data providers in the conclusion to identify relevant next steps necessary to consider AI solutions for finance and biodiversity.

# 1. Introduction

There is growing recognition among financial institutions, financial regulators and policy makers of the importance of addressing nature-related risks and opportunities. Over half of global GDP is heavily or moderately dependent on nature (World Economic Forum, 2020). Meanwhile, the World Wildlife Fund's (WWF) 2022 Living Planet report found that wildlife populations decreased by an average of 69% between 1970 and 2018, highlighting the magnitude of the biodiversity crisis (WWF, 2022). Furthermore, the Network for Greening the Financial System (NGFS) published an occasional paper declaring nature loss a material risk to financial stability, falling within the remit of central banks for management (NGFS, 2022; TNFD, 2023b). Demonstrating global consensus for action, the Kunming-Montreal Global Biodiversity Framework (GBF) adopted at the UN Convention on Biological Diversity's (CBD) COP15 in December 2022 outlines ambitious goals for 2030 to combat accelerating nature degradation (CBD, 2022). The catastrophic consequences of nature degradation at accelerating rates has never been clearer, and there have been calls for systemic-level change and engagement from the financial system to prevent major ecological collapse (Flammer et al., 2023; WWF, 2022).

Despite the increasing recognition of nature-related risks, previously little concrete guidance was available for financial institutions on identifying, managing or reporting their dependence or impact on nature. The Taskforce on Nature-related Financial Disclosures (TNFD) was convened to provide guidance for businesses and financial institutions on handling these nature-related risks and opportunities, releasing their final recommendations in September 2023 (TNFD, 2023b).[1] One aspect of the TNFD framework's theory of change is that with greater information on the nature-related impacts and dependencies of firms, investors can make more informed decisions and can reallocate their capital accordingly. While the recommendations outline the data and tools available to assess and disclose nature-related risks and opportunities, they do not include explicit recommendations for nature-related data or metrics useful for financial decision-making (TNFD, 2023b). Assessing nature-related risk is complex and requires location-specific considerations, which provide unique data requirements and challenges for financial institutions (TNFD, 2023c). To make informed and impactful financial decisions that align with nature-based outcomes and support investments with positive environmental impacts, it is essential to have the capabilities to transform the wealth of data into decision-useful metrics that capture broad geographic areas and granular information.

There are two key challenges with using data to determine investor impact on nature. Firstly, there is a preponderance of relevant heterogenous data in the form of, for example, corporate controversy text reports, disclosures, online environmental datasets, satellite imagery to name a few. Making sense of all this data at scale requires automation. Furthermore, it is a challenge to both express the complex interoperability between all organisations in the value chain and determine the impacts and dependencies they individually have on nature. The dual problem of scaling data analytics and analysing complex systems can be addressed using artificial intelligence (AI). With its ability to combine datasets, filter irrelevant information, clarify relationships within data, simulate the value chain and predict outcomes of financial decisions, we hope to demonstrate where state-of-the-art AI can have impact in both filling data gaps and attributing impact of finance on nature. Our project builds on emergent conversations exploring the potential of machine learning and AI in facilitating better decision making in sustainable investment (Kumar et al., 2022; Musleh Al-Sartawi et al., 2022). We attempt to address issues such as plugging existing data gaps with discovered data, data estimation under uncertainty, time series analysis and (near) real-time updates. We aim to pull together diverse existing datasets to develop nature-related metrics that better support decision making by financial institutions.

In this project, which ran from February 2023 to March 2024, we sought to meet the following research objectives:

---

[1] In line with the TNFD recommendations, we have defined nature as referring "to the natural world, emphasising the diversity of living organisms, including people, and their interactions with each other and their environment" (TNFD, 2023d).

- Understand use cases for how financial institutions use nature-related data for decision-making across different financial sectors, asset classes, and geographies, including existing data uses.

- Explore the feasibility of tapping into 'discoverable data' (e.g. satellite data, disclosures, traditional and social media, Internet of Things, *in situ* sensors).

- Explore how AI technology can deliver a data solution to measure and present asset-level nature-related risks.

- Understand complex interrelationships between actors in the investment process and the supply chain.

- Understand how ESG scores can be derived from granular data and how these scores scale to organisational data in such a way that they are consistent with third party scores.

To this end, we developed a conceptual framework for how best to combine geospatial data with biodiversity-related datasets and other relevant data sources to generate decision-useful metrics. We investigated the potential of AI to efficiently process large volumes of raw data, datasets, images, and maps. Our resulting framework outlines a model to enable the attribution of nature-related impact to financial institutions and large-scale auditing of self-reported corporate data on nature action.

## 1.1 Current State of Biodiversity Monitoring and Connection to Finance

With the growing pressure for financial institutions to assess, manage and disclose their nature-related impacts and dependencies, financial institutions require guidance on best practice to do so. Many biodiversity tools have been developed, such as the Integrated Biodiversity Assessment Tool (IBAT), Biodiversity Intactness Index (BII) and MapBiomas (TNFD, 2023a). Other resources, such as Exploring Natural Capital Opportunities, Risks and Exposure (ENCORE), the Nature Risk Profile (NRP), WWF's Biodiversity Risk Filter and CDC Biodiversité's Global Biodiversity Score, can help connect various types of financial institutions and their investments to their associated biodiversity impact and dependency (TNFD, 2023a; U. N. Environment Programme, 2023). However, targeted and actionable data driven metrics on the nature-related impacts of investments necessary for effective risk management and reporting are missing. Furthermore, without regulation to mandate disclosing nature-related risk metrics, the deployment and disclosure of these tools and frameworks remain voluntary. Additionally, without asset-level data from companies and financial institutions, often these tools and frameworks tend to lack comprehensiveness and granularity, which undermines the effectiveness and efficiency of data-driven investment decision-making on nature risks and impact.

With little publicly available information on nature-related risks at the company or asset level, many investors rely on ESG scores - akin to a credit score - to integrate sustainability considerations into their decision-making. However, these do not sufficiently capture nature-related risks and opportunities (Berg et al., 2019; Hughes et al., 2021; Rossi et al., 2024). Currently, ESG scores have large discrepancies between data providers, owing to the combination of heavy reliance on inaccurate, incomplete and incomparable self-reported data, as well as divergent methodologies in weighting various ESG metrics and calculating the final ESG score (Berg et al., 2019). Importantly, for the purpose of our research, the current array of ESG metrics, ratings, and scores do not comprehensively incorporate or reflect nature-related dependencies, risks, impacts or actions (Rossi et al., 2024; Xin et al., 2023). Although ESG metrics are designed to capture the broader environmental, social and governance factors for companies and investments and some may contain nature components, without incorporating any robust nature-related considerations, ESG scores in their current form will not be helpful in aligning finance for nature (Xin et al., 2023).

## 1.2 Current State of Biodiversity Risk Assessment Methods Used by Financial Institutions

To transition from ESG scores to explicit nature-related metrics for integrating nature into financial decision-making, methodologies must capture asset-level and attributable nature-related risk and opportunity. However, tracking and measuring nature-related impacts and dependencies is a challenging task (see Box 1) and current methodologies have many limitations. Often, current tools identify and measure nature-related risk at the country- or sector-level, which fails to sufficiently capture detail to assess different business models, ecosystem states or supply chain composition (Tiago Reis et al., 2023; van Toor & van Oorschot, 2020). Many methodologies do not sufficiently cover ecosystem types, such as marine or freshwater ecosystems, or the various drivers of nature degradation, such as invasive species, and, therefore, might underestimate nature-related impact (Sanyé-Mengual et al., 2023). Furthermore, most methodologies and tools capture potential values for nature-related risks due to their lack of granularity or clarity on the connection between ecosystem service benefits and the financial system, which can cause scepticism in their applicability or accuracy for policy or regulation decision-making (Villa et al., 2014). Therefore, while the methodologies and tools currently available (as discussed in Section 1.1 Current State of Biodiversity Monitoring and Connection to Finance) are helpful in gaining a first look into the impacts and dependencies of the macroeconomy on nature, they cannot provide sufficiently granular or accurate metrics to effectively guide financial decision-making. Data and metrics need to reflect localised considerations at the asset-level to communicate the actual nature-related risks and opportunities on economic activities while being comparable.

> **Box 1. The Challenges of Tracking and Measuring Biodiversity Risks and Impact**
>
> The challenges of tracking and measuring biodiversity risks and impact are four-fold.
>
> 1) **Interdependencies and interconnectedness of nature degradation and biodiversity loss:** Biodiversity is embedded in location-specific, and often interconnected, ecosystems and changes in one aspect of an ecosystem can have cascading effects on others. Compound risks related to other environmental stressors and social issues adds to the challenge of isolating and measuring the financial risk exposure of a particular asset, firm, capital provider, financial system, and economy. Furthermore, biodiversity loss can create positive feedback loops, whereby small losses in biodiversity lead to further biodiversity vulnerabilities and potential system collapse, making measuring biodiversity risks and impacts challenging.
>
> 2) **Complex supply chains:** Biodiversity loss and risks are often found within long, complex, and sometimes untraceable supply chains, often crossing international borders multiple times. This adds to the challenge of effectively tracing and drawing clear linkages between investments and on-the-ground biodiversity impact.
>
> 3) **Lack of standardisation:** There is currently a lack of standardised methodologies and metrics for assessing biodiversity-related financial risks and opportunities, making it difficult for companies, investors, and other interested stakeholders to compare and benchmark biodiversity performance. Central banks and governments are responding to the financial risks of biodiversity loss but related regulations and policies are still rapidly evolving. Companies and investors may find it challenging to anticipate and incorporate potential changes in regulations that could impact biodiversity-related financial risks. Coordinating efforts and establishing a standardised approach to measure and track biodiversity-related financial risks on a global scale can be challenging.
>
> 4) **Valuation:** The highly diverse landscape of biodiversity measurement and tracking methodologies adds complication to assigning economic values to biodiversity risks, impacts,

> and opportunities. While efforts have been made to develop methods for valuing ecosystem services, the subjective nature of these valuations and the lack of consensus on pricing can hinder accurate financial risk assessment.

*Box 1. Challenges of Tracking and Measuring Biodiversity Risks and Impact*

## 1.3 Robust Data and Methodologies for Material Biodiversity Risk Measurement

### 1.3.1 Geospatial and Asset-level Data

In the face of these challenges, there have been calls for more robust, transparent and auditable nature-related metrics and data disclosure (Caldecott, McCarten, et al., 2022; Hughes et al., 2021; Rossi et al., 2024). As ecosystems can be profoundly impacted by localised risks, increasing granularity of models and data to make them usable at the asset-level is key to improving the robustness of nature risk assessment. Therefore, incorporating granular geospatial data, such as satellite imagery, that captures nature-risk and impact at the asset-level rather than relying on sectoral or country averages, is an excellent first step in improving the underlying data that informs ESG scoring and also nature-related risk and opportunity measurement (Rossi et al., 2024).

Spatial finance, the integration of geospatial data and analysis into financial theory and practice, provides a useful framework to integrate geospatial data into financial modelling and decision-making (Caldecott, McCarten, et al., 2022). Key benefits of geospatial information include timelines and assurability (Caldecott, McCarten, et al., 2022). Therefore, it can help provide investors with reliable information relating to actual rather than potential nature-related risk that is not reliant on self-disclosure (Berg et al., 2022; Caldecott, McCarten, et al., 2022). By integrating verifiable geospatial data, there could be a real opportunity to improve the comparability and transparency of nature data and metrics, and thus enhance the robustness of measurement (Rossi et al., 2024). In this sense, AI can provide a technique to link robust geospatial biodiversity data to E-scores and, therefore, contribute to the attribution of nature-related risk to the actions of companies, helping to bridge a current gap.

### 1.3.2 AI Applications and Modelling Approach

In our proposed approach, relationship diagrams were developed by illustrating the connections between *actors* in each use case in terms of flows of ecosystem services, nature degradation, financial information, capital and environmental regulation, among other factors. Probabilistic expressions were then derived to describe the relationships outlined in the relational diagrams. The probabilistic model allows epistemic uncertainty in these relationships to be expressed explicitly and is defined in such a way that value chain interoperability can be broken up into individual probabilistic factors. Each factor models the behaviour of a single organisation, agent or resource within the supply-chain, thus allowing a multiple disciplinary approach to model building by experts in respective fields. From the probabilistic expressions, the data required to set initial conditions and provide information to the model can be determined. These agent-based modelling approaches are transparent, flexible and handle uncertainty well, making them excellent candidates to model the financial system and complex socio-ecological systems, as well as their connections (Farmer et al., 2015). With interconnected systems as complex and uncertain as nature and finance, the Bayesian modelling approach is particularly useful as it can identify policy-relevant scenarios through a data-first pipeline with fewer arbitrarily determined assumptions or hidden uncertainties than current methodologies (Farmer et al., 2015).

Our proposed models are conceptually developed for both use cases. However, they have not been tested nor validated, as this was beyond the scope of the current project.

## 1.4 Use Case Selection

In this report we focus on two use cases to demonstrate how AI could contribute to integrating robust biodiversity metrics into ESG scores and better financial decision-making. Our two use cases are:

- Bank Lending to the Brazilian Cattle Ranching sector
- Asset Manager Bond holding in the UK Water Utility sector

Through the use cases, we demonstrate AI solutions for different aspects of integrating finance and nature, namely a top-down greening finance example and a bottom-up financing green example. The first use case focuses on bank lending to the Brazilian cattle ranching industry and represents a **greening finance** use case. The second use case focuses on an asset manager holding corporate bonds in a UK-based water utility company and is an example of **financing green**. Furthermore, both sectors have high reliance on nature and also a high impact on nature from their activities, indicating that nature-related risks are likely to be material from a financial or investor perspective.

# 2. Use Case 1: Brazilian Cattle Ranching Sector

Our first use case focuses on the cattle ranching sector in Brazil, particularly within the legal Amazon[2]. This proved to be a useful use case due to the strong combination of:

- High risks to biodiversity, driven by the extensive land conversion linked to cattle ranching;
- The biodiversity importance of the region;
- Clear links to financial institutions (who provide loans to farmers and slaughterhouses), and
- A strong legal framework in the shape of Brazil's Forest Code (Chiavari & Leme Lopes, 2015).

There is also a strong body of existing literature and data in this space on which to build our AI model.

## 2.1 Introduction

The Brazilian agricultural sector, specifically the beef and soy sectors, have been major drivers of ecosystem conversion and degradation in the Amazon and Cerrado biomes in recent years (Pendrill et al., 2022; Tiago Reis et al., 2023). The resulting direct, negative, and compound impact on biodiversity and global climate action has led to pressure building on financial institutions who are directly or indirectly funding this destruction through financing the agricultural sector. Reputational risks are growing for such institutions, as are the nature-related physical risks arising from future losses of ecosystem services, which could then translate into financial risks. The nature-related financial risks and reputational risks have the potential to be particularly pronounced for this use case as the Amazon is one of the most biodiverse biomes in the world and its degradation is the focus of significant publicity.

Brazilian banks are relatively highly dependent on ecosystem services, exposing them to risk if nature is degraded to a point where these ecosystem services are no longer delivered. For example, the World Bank recently found that 46% of Brazilian banks' non-financial corporate loan portfolio is concentrated in sectors highly or very highly dependent on one or more ecosystem services (Calice et al., 2021). Moreover, 15% of Brazilian banks' corporate loan portfolio has exposure to firms potentially operating in protected areas (Calice et al., 2021). However, many financial institutions struggle to understand their impacts on nature, and how this relates to material financial or non-financial risks, as the many existing data sources are not in a decision-ready format. In this project, we look at ways of improving the availability of decision-grade nature data to support better understanding of such risks.

### 2.1.1 Brazil's Forest Code

A key legal mechanism enabling the integration of nature-related considerations into the Brazilian banking system is the Forest Code, formally known as the Native Vegetation Protection Law, No. 12,651/2012 (WWF, 2016). This legislation defines a comprehensive set of requirements that private landowners must comply with to ensure a balance between land use and the protection of natural ecosystems and the services they provide. However, issues related to enforcement and transparency pose challenges to the Forest Code's effective implementation. Three key aspects of the Forest Code relevant to our use case are 1) the state licence requirement for deforestation activities, 2) Permanent

---

[2] Legal Amazon is a division defined by the Law to differentiate it from the biome, as well as the International Amazon. Legal Amazon refers to the entire Amazon biome, as well as parts of the Cerrado and Pantanal biomes. This region was defined as a region that can articulate activities aiming at its socioeconomic development, and is conformed by nine states: Amazônia, Acre, Rondônia, Roraima, Pará, Maranhão, Amapá, Tocantins, and Mato Grosso (Instituto Brasileiro de Geografia e Estatística, 2014).

Preservation Areas and minimum Legal Reserve requirements, and 3) the establishment of the Rural Environmental Registry (Cadastro Ambiental Rural - CAR):

1) *The state licence requirement for deforestation activities.*
   The Forest Code establishes that landowners must obtain a licence before any conversion of natural ecosystems. However, public data on licences is not always readily available or comparable, making the legality assessment challenging (Vasconcelos et al., 2023).

2) *Permanent Preservation Areas and minimum Legal Reserve requirements.*
   The legislation defines Permanent Preservation Areas (*Áreas de Preservação Permanente*, APP) as areas that must be left intact because of their critical contribution to preserving essential ecosystem functions, such as regulating weather and hydrological cycles and ensuring a steady water supply. The law specifies areas that are considered to be APPs, which include banks of rivers, springs and lakes, mangroves, *vereda* wetlands, hilltops, steep slopes and sandbanks.

   Beyond APPs, the Forest Code also requires that a proportion of land on a rural property is set aside for conservation and sustainable use of natural resources. This proportion is known as Legal Reserve (*Reserva Legal*) and varies depending on the biome. It is set at 80% in the Amazon rainforest, 20% in the Pantanal, and 35% in the Cerrado (Ro, 2023).

3) *The establishment of the Rural Environmental Registry (Cadastro Ambiental Rural - CAR).*
   Finally, the Forest Code mandated the establishment of the Rural Environmental Registry (*Cadastro Ambiental Rural*, CAR). All rural properties are required to be registered in this government-run registration system (Sistema de Cadastro Ambiental Rural, n.d.). Property owners are responsible for inputting the relevant information regarding their property into the registry, including ownership details, APPs, Legal Reserves and georeferenced information on the property boundaries (Sistema de Cadastro Ambiental Rural, n.d.). The CAR is a powerful tool for integrating nature considerations into the financial decision-making and portfolio management by Brazilian banks due to its wealth of spatially explicit environmental information (Soares-Filho et al., 2014), and we aligned our AI model with this key dataset.

## 2.1.2 Brazilian Beef Industry

In 2020, the Brazilian beef industry cleared 291, 955 hectares of forest in the Amazonian region for cattle pastureland (Trase, 2023). Despite many of the largest international banks having commitments against investing in deforestation, many continue to provide credit for and purchase shares in meatpackers who have been linked to deforestation activities, which exposes them to reputational and litigation risks (Moye, 2022). As a result, many banks and other financial institutions are facing increasing pressure due to increasing investigative reporting from NGOs and other activists, such as Global Witness' 2022 'Cash Cow' report which named and shamed major international banks and asset managers for their continued financing of JBS and other major Brazilian meatpackers despite mounting evidence of illegal deforestation in their supply chains (Moye, 2022).

The three largest meatpackers in Brazil – JBS, Marfrig and Minerva – have implemented deforestation monitoring systems within their value chains in the Amazonian biome. However, these systems have not been effective in eradicating deforestation exposure in their supply chains due to lack of traceability of the supply chains, and deforestation occurring in indirect suppliers who are not properly traced. According to [Trase](#), an agricultural deforestation-focussed research initiative, JBS was exposed to 230,000 hectares of cattle deforestation in 2020 through its supply chain, while Marfrig was exposed to 110,000 hectares, and Minerva 91,000 hectares (Trase, 2023). The large Brazilian meat packers have faced significant controversy in recent years for their environmental degradation as well as human rights abuses. Global Witness notes that JBS source directly from 144 ranches in the Amazon state of Pará that fail to comply with legal agreements. Further, JBS also failed to monitor an additional 470 ranches further up its supply chain, containing an estimated 40,000 football pitches of illegal Amazon clearance (Moye, 2022).

As the public interest in beef-driven illegal deforestation in the Amazon continues, both Brazilian and international banks will face increased reputational and transition risks, which has the potential to increase their cost of capital or revoke their social licence to operate. In Brazil, banks are expected to screen borrowers for compliance with the Forest Code. Although civil society has raised concerns over weak and uneven implementation of the Forest Code (Chiavari & Leme Lopes, 2020), lending to non-compliant farms, abattoirs, and food processing companies nonetheless poses regulatory risks. Beyond compliance reasons, banks should manage the deforestation risk of their loan portfolio as part of their climate and biodiversity commitments, as well as to safeguard their reputations.

In this use case, we built a model of where each abattoir likely sourced their beef, which would allow financial institutions to access their nature-related risk exposure to lending to this sector.

### 2.1.3 Financial Context: Focus on Sustainability-Linked Loans

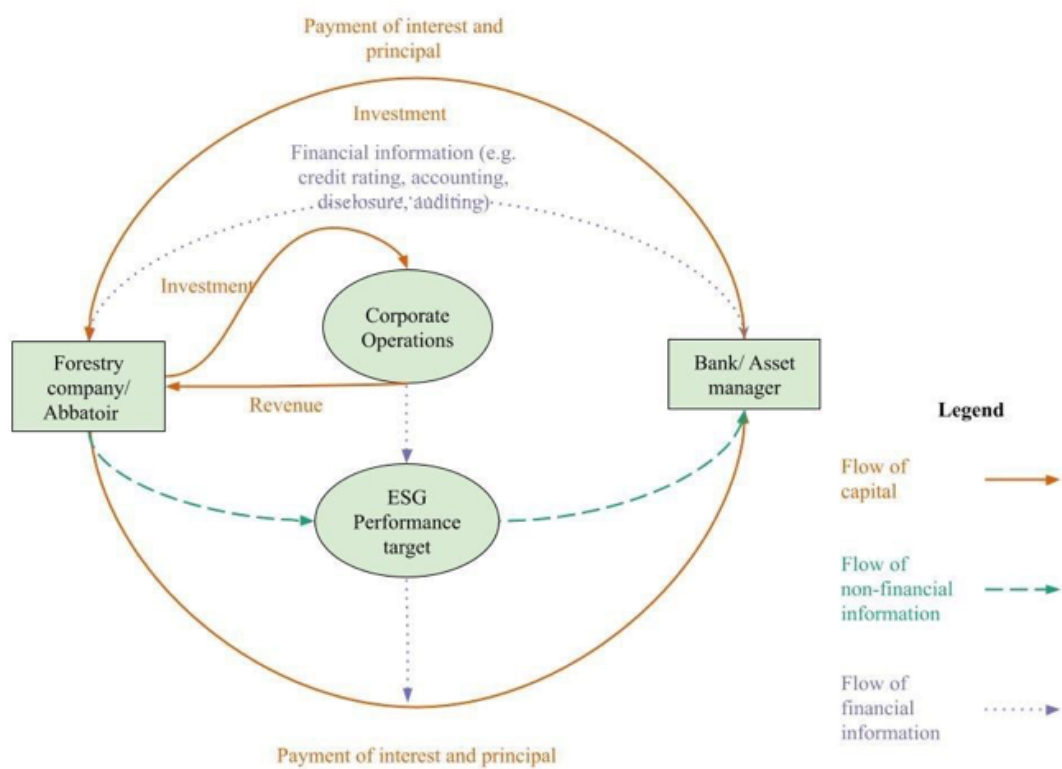

*Figure 1. A diagram describing the flow of capital, non-financial information and financial information in the Brazil cattle industry. Source Credit: Author's own*

As the Brazilian beef sector is heavily reliant on loans, we suggest that sustainability-linked loans (SLLs, see Figure 1) may provide a solution to integrate material biodiversity risks, such as compliance and reputational risk, into lending decisions for the sector. This instrument was formally codified in 2019 to support sustainable transitions of corporates (ICMA, 2020; LMA, 2019). Unlike some green bonds and green loans where the proceeds of the debt instrument are ring-fenced for specific eligible 'green' assets (e.g. a renewable energy plant, or a green building), the proceeds of an SLL can be deployed for general corporate purposes. Prior to pricing the bond or borrowing an SLL, the borrower commits to a series of ambitious sustainability performance targets that are material to their business. They also commit to a 'step up' clause - if the company fails to meet its sustainability performance targets it will have to pay a higher interest rate as a penalty - and/or a 'step down' clause - if the company successfully meet its sustainability performance targets it will pay a lower interest rate as a reward. These clauses act as additional financial incentives to meet the pre-agreed sustainability performance targets. The financial case for putting the step up and step down clauses in place is that companies which successfully meet their sustainability-performance targets are exposed to fewer environmental, social, and governance risks, which should reflect positively on their ability to pay back their debt. The proficient handling of nature and deforestation risks holds paramount significance for the beef sector in mitigating physical, reputational, and regulatory risks. Incorporating these considerations into loan agreements can offer financial incentives for companies operating along the beef supply chain, ranging from farms and abattoirs to meat processors and upstream buyers to enhance their management of nature dependencies, risks, and impacts.

Although sustainability-linked instruments have been most popularly used for reducing corporate carbon intensity and emissions, they are increasingly adopted to support corporates in achieving their biodiversity targets, although to different levels of success. For example, in 2019, Marfrig issued the first sustainable transition bond in Brazil worth USD 500 million (Marfrig, 2019). The 10-year bond supports their acquisition of cattle from suppliers with best practice traceability for preventing deforestation and ensuring animal welfare and with a strong record on labour and human rights (Marfrig, 2019). Later in 2021, Marfrig secured a USD 30 million sustainability-linked loan from Netherlands-based impact investor &Green Fund to achieve full supply chain traceability, promoting technical assistance to support suppliers' ability to meet compliance requirements, and other environmental and social standards that goes beyond the Forest Code (IDH Sustainable Trade Initiative, n.d.). In the same year, another leading meat producer, JBS, raised $3.2 billion in the U.S. market after committing to reduce its emissions by 30%, compared with 2019's baseline, in line with pursuing the goal of becoming net-zero in 2040. It is worth noting that JBS's bond has faced criticisms for failing to include deforestation and cattle enteric fermentation (MightyEarth, 2023), suggesting that accountability mechanisms of sustainability-linked debt instruments require more robustness checks.

The issuance volumes of SLLs have decreased significantly in 2023, as the result of growing investor concerns and civil society criticism over the lack of credibility of the sustainability performance targets, which are heavily reliant on unreliable ESG scores and ratings. Our model could offer a more robust, data-based, and bespoke indicator of the ESG performance of firms.

## 2.2 Methodology

There were two phases to this work: first, research on defining the use case through desk-based research and interviews with relevant financial institutions, and second, developing the AI model to fit this use case. We then refined the AI model by further conversations and research.

To start, we conducted desk-based research on existing climate and biodiversity policies of the top ten Brazilian banks to understand their key biodiversity priorities, data use and sources, and any existing methodologies. This revealed a lack of detailed, quantitative public information disclosure on these topics.

To dig deeper we reached out to a range of Brazilian financial institutions and conducted interviews to understand how they are currently considering biodiversity and nature risks when it comes to investment in the agricultural sector. We also engaged with existing biodiversity data providers, including Trase and BVRio, to deep dive into potential machine learning use cases to enhance the scope and efficiency of data collection and analysis. Additionally, we identified robust sources of open-access data which are relevant to this use case, and which could be integrated into an AI model.

Once we had a good overview of the challenges in this space, we started to draft the AI model. Here we used a combination of literature reviews and innovative thinking to develop our model outline, drawing on existing research in this space. A key concern was building a model that satisfied the requirements of the financial, ecological, logistic and commercial agencies involved in the supply chain.

Finally, we reached out to some of the stakeholders that we had initially interviewed to get feedback on the model and its potential use within banks and other financial institutions. We also discussed the model with the TNFD, in the context of the potential to validate future disclosures by companies in line with the TNFD framework (see Section 2.4.11 Feedback Interviews with Relevant Organisations for details of this discussion).

## 2.3 Results

### 2.3.1 Needs from Financial Institutions

To scope out the needs of financial institutions in this space, and understand how they currently assess risks, we conducted five scoping interviews.

The following points were common across two or more of the interviews:
- Biodiversity is not considered as large a material risk as climate change. Financial institutions recognise that biodiversity loss is becoming a progressively more important risk, but there is still no clear understanding of the business case for better integration of biodiversity information into decision making.
    - Deforestation, by comparison, was seen by several interviewees as a material issue. However, this again tended to relate to the importance of deforestation in the climate agenda.
    - There are, however, increasing litigation risks for financial institutions in this space, and a clear legal framework in Brazil in the form of the Forest Code. Therefore, relating nature risks to litigation risks may be a good avenue to demonstrate materiality for financial decision making.
        → *Suggest developing a metric which connects nature risk to regulatory risk using the Forest Code as a key link.*

- Data availability was not seen as an issue - but centralising and making that data decision ready was identified as a challenge. Land cover and land use change data for Brazil is particularly abundant, as it is a highly studied country - however several of the financial institutions we spoke to highlighted how difficult it was to pull this data, in its current form, into their decision-making processes, where nature risk was just one aspect under consideration.
    → *Anything we produce should focus on drawing together existing data layers and making them accessible for decision makers within banks.*

- Banks often collect CAR data as part of their compliance processes, however there are known issues with the accuracy of reported information. (The larger asset managers which we interviewed did not collect this information as they did not deal directly with farmers, but rather with other, larger, agribusinesses). Other data gaps identified mostly relate to government systems or corporate reporting.
    → *Cost-effective and automated ways of verifying the accuracy of CAR data would be a valuable addition in this space. This would include, for example, the potential use of AI to look at CAR data boundaries and compare these to satellite imagery to assess if they are likely to reflect true boundaries on the ground.*

- The larger banks we spoke to do not finance individual farmers, but rather offer loans to larger aggregators in the supply chain such as abattoirs and meatpackers.

> *Therefore, working at the level of the CAR unit is not so useful, but finding a way to aggregate nature risk up the supply chain, e.g. to the abattoir or meatpacker, would be more helpful.*

Some further takeaways from specific interviews are as follows:
- 'Bank A' (an international bank with operations in Brazil, who asked to stay anonymous)
    - Does not directly finance agricultural suppliers - is limited in scope to lending to global corporations. They find that supplier management in the commodity space is difficult due to quick changes in supplier lists. Would find a nation-wide traceability system and a more comprehensive monitoring of indirect supply chains very helpful to assess their risks.
        > *We propose using AI tools and existing research on cattle sourcing models in the vicinity of abattoirs to develop a probabilistic model of cattle supply chains.*

- JGP Asset Management
    - Stated that a key issue currently is not in data collection but more so in centralising and interpreting the data that is already available. Currently their ESG team has to reach out to primary and secondary data sources to compile them together.
        > *Further evidence to support a model which compiles existing data sources into an actionable metric.*
    - Also supporting meatpacking companies to reach deforestation-free targets by 2025 through the issuance of sustainability-linked bonds.
        > *We will continue to consider sustainability-linked bonds as a finance mechanism to apply this metric.*

- Santander Asset Management
    - Currently combine different data sources, looking for information from companies' reports and ESG data providers. However, these are not fully reliable and local analysis is sometimes necessary to ensure compliance and accuracy of reported information. Biodiversity risk from scientific literature still has to be effectively translated into financial risk in order to influence their investment decisions.
        > *Further evidence to support a model which compiles existing data sources into an actionable metric.*

Through our proposed models and potential AI implementations, we have focussed on addressing the challenges highlighted above.

### 2.3.2 Nature Risk Profile: A Framework to Calculate Nature-Related Impacts

We used the [Nature Risk Profile (NRP) methodology](#), developed by UNEP-WCMC in collaboration with S&P Global, to structure the drawing together of data layers for this use case. This methodology draws heavily on the principles outlined by TNFD and is a methodology for profiling nature-related risks associated with location-specific business activities.

The NRP methodology rests on two core building blocks for profiling nature-related risks: dependencies on nature and impacts on nature. **We focused our investigations on the impacts methodology**, which is summarised in Figure 2 below. This provides an approach for financial institutions and companies to calculate nature impact-related risk at either the individual asset level or the company level by "estimating footprints for magnitude of impact associated with business activities and analysing the metrics for location significance where these impacts occur" (U. N. Environment Programme, 2023). The methodology provides a structure in which to combine different sources of nature data using the key building blocks of magnitude and significance, which helps address the issue we heard from financial institutions of there being too many different data sources, and not knowing how to make them decision-ready.

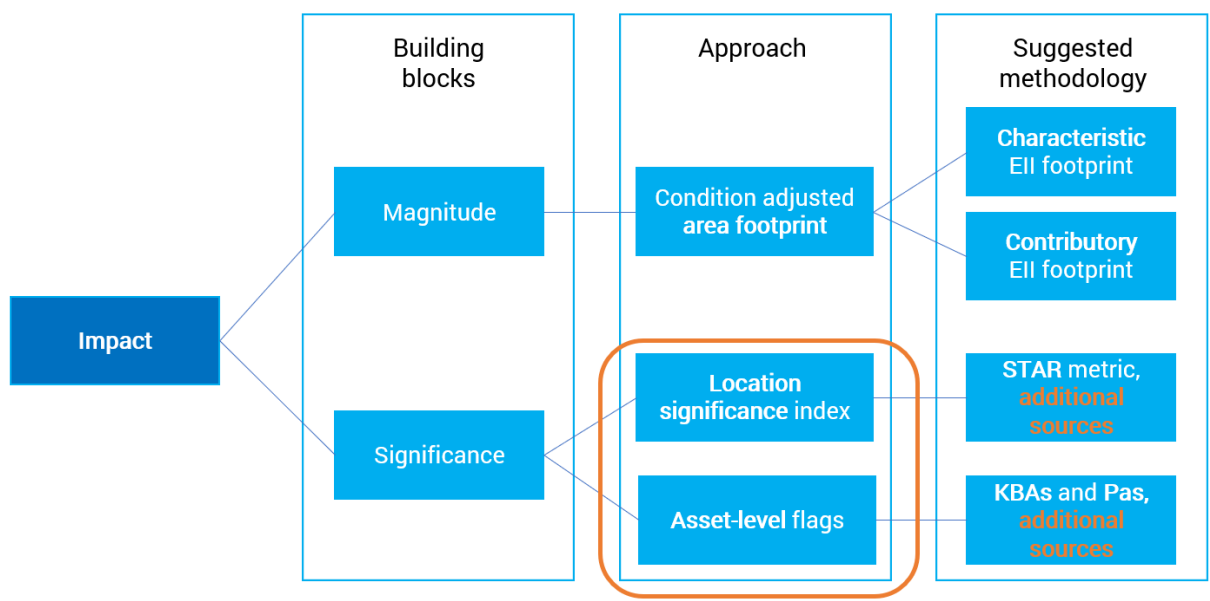

*Figure 2. Nature Risk Profile (NRP) impacts methodology.*

We suggest applying this methodology to the Cadastro Ambiental Rural (CAR) data in Brazil, as a way to assess the nature impact within each CAR unit.

Key to the NRP methodology is the concept of a condition-adjusted area footprint which involves quantifying the extent of ecosystem coverage in an area of interest and then reducing this total extent by a factor representing its condition compared to an intact reference state. This concept can also be applied to a business activity (such as agriculture), with the total area of land occupied by a business activity adjusted for the degree to which the condition is reduced. The Ecosystem Integrity Index is suggested as the best practice metric for calculating the 'remaining condition' – we suggest calculating this for each farm as registered in the CAR database.[3]

The second key pillar of the NRP approach is the use of a location significance index and asset-level flags to add a 'significance' element to complement the magnitude of impacts calculated using condition-adjusted area footprints. This allows the relative significance of the ecosystems impacted to be taken into account (significance includes aspects such as biodiversity richness, provisioning of ecosystem services, and species extinction risk). For the local significance index and asset level flags, we look to local datasets to add better local context and complement global data layers, as highlighted in orange in Figure 2 and expanded upon in Section 2.3.3.

### 2.3.3 Examples of Existing Data Sources

There are many datasets on the state of biodiversity and nature in Brazil. However, for the most part, this data is not in a form which is able to directly inform decisions made by financial institutions. This is partly because linking environmental data through to the impacts a company is having on nature requires a range of skill sets - including a thorough understanding of the environmental datasets themselves – an expertise most FIs currently lack. Additionally, many of the datasets are not high enough resolution to be able to assess asset-level impacts or are not updated frequently enough to track how the actions of an FI are affecting nature within their portfolio.

---

[3] The Ecosystem Integrity Index (EII) developed by UNEP-WCMC aims to help businesses and financial institutions in their understanding and reporting of biodiversity measurements. The EII covers all the components of an ecosystem (structure, composition, and function) in a single index.

In Table 1 below we outline a range of existing data sources which we identified as most relevant to this use case, and which could add greater granularity when integrated into the NRP methodology. By bringing these datasets into the NRP methodology, we are able to draw out a more actionable level of information from them - namely, what is the nature-related impact generated within each CAR.

*Table 1. Survey of relevant environmental datasets and methodologies*

| Data Source | Description | Project Relevance |
|---|---|---|
| **[Cadastro Ambiental Rural](#)** *(translates as 'Environmental Rural Registry')* | Brazil's Forest Code established a mandatory and georeferenced registry of farms aimed at advancing the implementation of its natural ecosystem protection and restoration requirements. CAR data can be accessed through a centralized portal maintained by the federal government, but it is collected and implemented at the state level. Not all data makes it to federal level portal, with some held in state level systems. | Access to >6m properties (over 544m ha), which provides us with the spatial element to link analysis to. Many financial institutions have access to the CAR numbers of the farms they are currently investing in. **Issue:** The accuracy of data provided by CAR has been challenged, as data is self-declared and not always verified ([Chain Reaction Research, 2022](#)), or updated. |
| **[World Database on Protected Areas (WDPA)](#)** | The World Database on Protected Areas (WDPA) is a joint project between UN Environment Programme and the International Union for Conservation of Nature (IUCN), managed by UNEP-WCMC. Data for the WDPA is collected from international convention secretariats, governments, and collaborating NGOs. The WDPA uses the IUCN definition of a protected area as the main criteria for entries included in the database.<br><br>The dataset can be found on the Integrated Biodiversity Assessment Tool (IBAT) and is available for commercial use. The Brazil country profile is found [here](#). | Protected area data is required for the 'significance' element of the NRP methodology. This global dataset can be complemented by other local datasets such as local maps of indigenous lands or other areas of biodiversity significance. |
| **[Key Biodiversity Areas (KBA)](#)** | KBAs represent the most important sites for biodiversity conservation worldwide, and are identified nationally using a Global Standard from the International Union for the Conservation of Nature (IUCN).<br><br>The dataset can be found on the Integrated Biodiversity Assessment Tool (IBAT) and is available for commercial use. The Brazil country profile is found [here](#). | Key Biodiversity Area data is required for the 'significance' element of the NRP methodology. This global dataset can be complemented by other local datasets such as local maps of indigenous lands or other areas of biodiversity significance. |

| [Species Threat Abatement and Restoration Metric (STAR)](#) | The Species Threat Abatement and Restoration Metric (STAR) allows quantification of the potential contributions that species threat abatement and restoration activities offer towards reducing extinction risk across the world. STAR is calculated from data on the distribution, threats, and extinction risk of threatened species derived from the IUCN Red List of Threatened Species™.<br><br>An Estimated STAR score is available through IBAT Reports. This global data layer can then be updated to a Calibrated STAR score that confirms the presence of threats and species at a site, establishing a baseline against which conservation management can be planned and targets set. | STAR data is required for the 'significance' element of the NRP methodology.<br>***Issue:*** Estimated STAR has a 5km$^2$ resolution, so is not granular enough to offer much distinction at the level of a farm or asset. It is instead intended to highlight areas within a landscape which have a higher level of biodiversity importance. Additionally, there are limitations related to how frequently Estimated STAR is updated, as it is directly derived from the IUCN Red List data. Calibrated STAR scores should be used for more granular baseline assessment. |
|---|---|---|
| [Ecosystem Integrity Index (EII)](#) | The EII metric represents the integrity of terrestrial ecosystems globally at 1km$^2$ resolution. It is formed of three components - structure, composition, and function - and measured against a natural (current potential) baseline on a scale of 0 to 1. The EII is currently accessible in [pre-print form](#), and when fully published the global EII layer will be included in tools such as the [UN Biodiversity Lab](#). | EII is used to calculate the magnitude of the nature impact in the NRP methodology, through developing an adjusted area footprint. |
| [Mapbiomas](#) | High resolution maps with historical data on land use and land cover in Brazil, allowing greater visibility of which areas/regions are most at risk of deforestation. | Provides data on recent land use change – recent deforestation or natural ecosystem conversion could be used as an asset level flag. |
| [Instituto Brasileiro do Meio Ambiente e dos Recursos Naturais Renováveis (IBAMA)](#) | Open-source database of fines and embargoed areas, with shapefiles. The data comes from the federal environmental agency (IBAMA) and the system is frequently updated (automatically as fines are applied and areas are embargoed). State environmental data can be used to complement the federal information since agencies also hold data on fines from their own enforcement activities, but in this case the data is not always easily accessible. | Could be used to inform asset level significance flags in NRP. Some banks use this data to look at risks associated with potential investment areas. |

| | IBGE collects and releases their Pesquisa da Pecuária Municipal (PPM), which provides annual estimates of herd size by municipality (Trase, 2023b). | This can provide an estimate of cattle numbers in the region and any changes, which can be helpful for estimating land conversion due to cattle ranching. |
|---|---|---|
| [WWF Water Risk Filter](#) | Provides a global map of water-related physical, regulatory and reputational risk that helps investors screen for potential water risk exposure based on the location and sector of the investment. The tool also includes built-in scenarios for water risk that allows investors to explore the potential impact of water risk on their portfolio or investment (WWF, 2021). | Water-related risk could be used as an additional data point to provide location significance. |
| [Animal Transit Permit (GTA)](#) | The Animal Transit Permit is a mandatory record system that tracks cattle records and cattle transportation. When a cattle lot arrives at a meat processor, it will be accompanied by a GTA detailing the origin of the lot. However, the GTA will not detail prior farms that the lot may have passed through on its way to the final lot and the meat processor, which limits its usefulness as a traceability tool. Furthermore, the GTA is fully publicly or digitally available (Chain Reaction Research, 2018; Proforest, 2017). | The permits help link abattoirs to beef suppliers, which is used in our proposed model for supply chain tracing. |
| [Trase](#) | Estimates and attributes deforestation to agricultural commodity supply chains and their funders for various geographies by importer country and producing country.<br><br>For Brazilian Beef, Trase primarily uses abattoir and export data from Brazilian Ministry of Agriculture, Ministry of the Environment (IBGE) to estimate deforestation in supply chains. On their dashboard, they calculate hectares of deforestation exposure of large meat processors, importer countries and municipalities in the Amazonia and Cerrado biomes. | Supply chain data for direct supply chains.<br><br>***Issues:*** Based on outdated Government statistics (newest information from 2019). |

### 2.3.4 Potential Data Gaps and Issues Identified

In addition to the issues raised by banks, an assessment of literature, grey literature and existing datasets highlighted the following points, which we have also integrated into our development of an AI model:

- There is currently a **lack of indirect supply chain data** to give banks visibility over the whole value chain. This information would help tracking the nature-related impact of companies and suppliers further upstream in the supply chain and would help capture the indirect impacts that banks investments have on nature.

- Whilst we can work at the level of CAR data on the ground for producers/farmers, this does not help if investors are funding companies in the supply chain, who are removed from the immediate point of agricultural production. Supply chain information would help us track the nature impact of, for example, an abattoir or a trader who might be a bank's client.

- Assumptions of how cattle move through this chain are likely to be non-trivial: cattle laundering is a major issue in the Brazilian beef sector.

- Trase have done a lot of good work in tracing direct supply chains for the beef sector in Brazil, and exports from there, but they have not fully captured indirect supply chains. However, the Beef on Track initiative and Para state's Cattle Integrity and Development Program are looking to help fill these crucial data gaps (Andre Vasconcelos et al., 2024).

- **Inconsistency of temporal and spatial resolution of datasets** is likely to be an issue.

- Additionally, some datasets may not be granular enough for our needs. For example, the Ecosystem Integrity Index (EII) is at 1km² resolution, which may be too large to be useful at a farm level. There is a need for more granular datasets.

- It would be ideal to have more local, granular data also on the extent and quality of ecosystems, for example the state of ecosystem or species richness.

## 2.4 Proposed Model

Our model takes a Bayesian approach, which captures the relationships between the actors in the supply chain. The Bayesian approach uses Bayes theory to explicitly encode probabilistic relationships between the actors and also the model parameters. Probability is important as it allows us to express precisely what we know and the level of certainty in what we know. Even weak information can be expressed using this approach and this is a critical requirement for nature finance applications for which many parts in the value chain are only partly or weakly known. This approach allows the information to be inserted into the model as it becomes available around certain assumptions rather than building hard assumptions on the data. The Bayesian approach is very general and supports the inference of model parameters from data, the fusion of different data sources, time series analysis, prediction and many more tools (Bishop, 2006). Bayesian models are helpful for us because they can be visually represented as a network of actors (or "agents") and their connections (Villa et al., 2014). This proposed model was conceptually developed but not tested or validated, which was beyond the scope of the current project.

The actors, and relationships between them, are shown in Figure 3, where we formalise our working knowledge on this use case in order to then take it forward into an AI model.

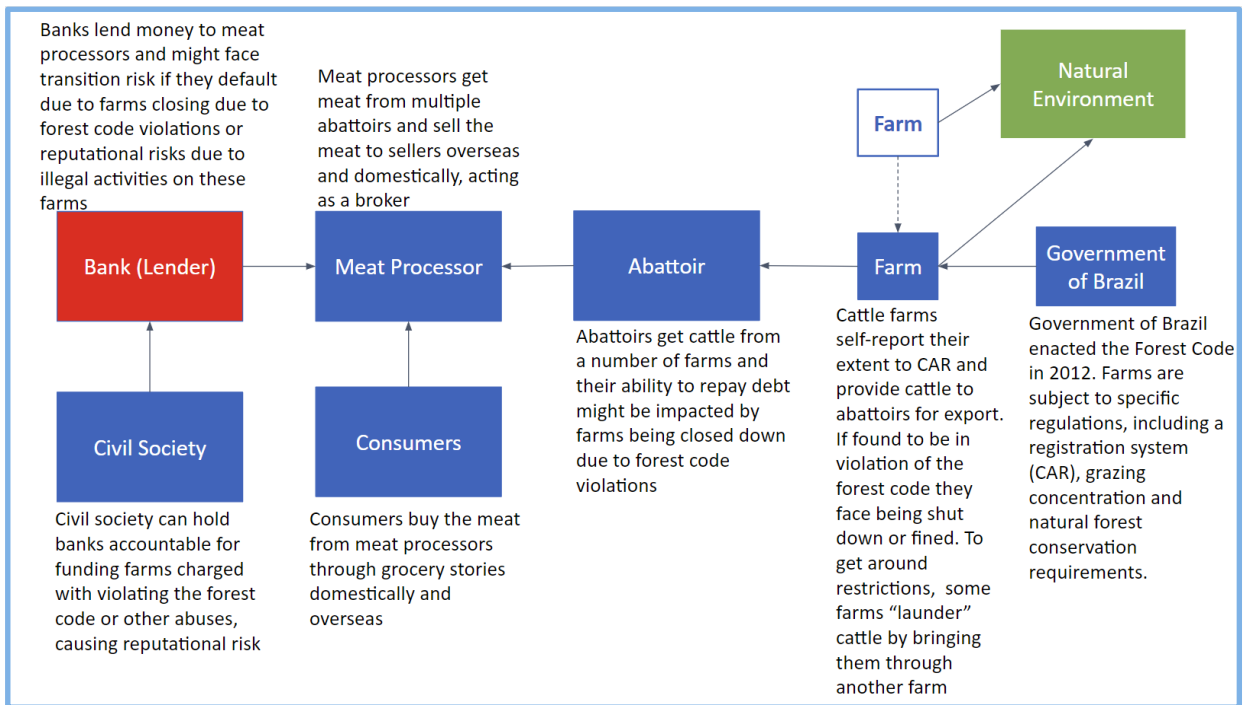

Figure 3. Relationship diagram for actors in the beef supply chain.

In the use case, as seen in Figure 3, the Brazilian banks provide debt to meat processors, such as JBS, which sources beef from a series of abattoirs, which in turn source cattle from a number of different farms. The farms have a direct impact on the natural environment through deforestation, overgrazing, water intensive practices, pollution from fertilisers, among other impacts. The Government of Brazil regulates the farms through the Forest Code. To avoid these regulations, some farms "launder" their cattle through a second farm before reaching the abattoir to allow them to graze more cattle than permitted and avoid detection by enforcement agencies. A simplified example of how that looks like is in Figure 4 below, developed by (Proforest, 2017). The range of tools currently available to monitor and trace cattle are only limited to direct suppliers.

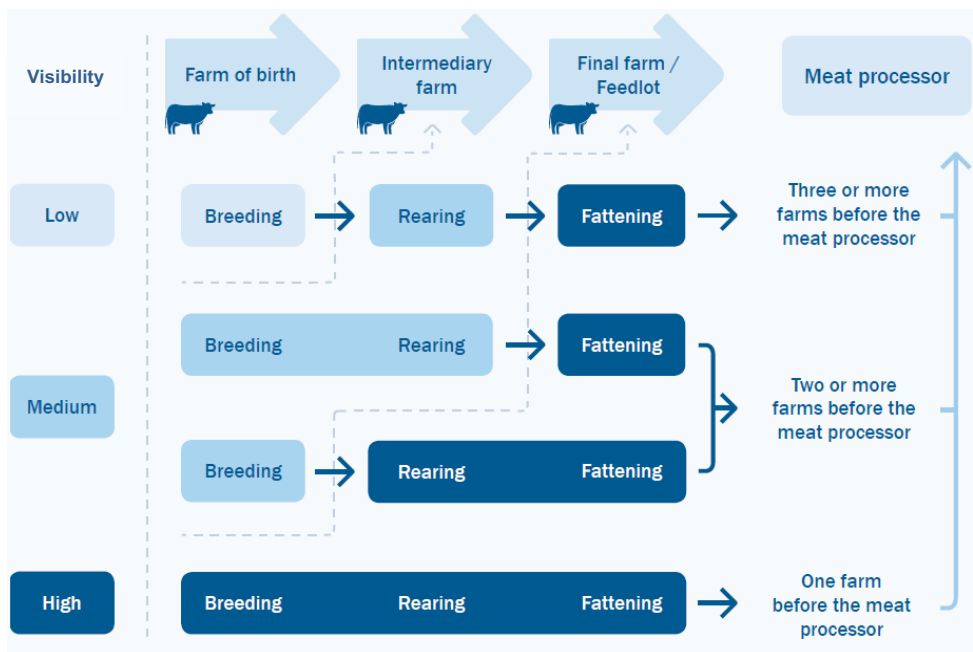

Figure 4. Levels of traceability of the different stages of cattle production (Proforest, 2017)

Farms found to be in violation of the Forest Code can be fined and embargoed by the Government and are not eligible to receive loans from Brazilian banks. Some meat processors have consistently faced controversy due to their continued purchasing of beef from farms participating in illegal activities from deforestation to cattle "laundering". To limit reputational and regulatory risk, banks could use the proposed model to link loans to sustainability criteria by predicting the probability farms in the supply chain violating the Forest Code and creating a corresponding "ESG" score for each farm in the supply chain. We explore this approach later in this section.

### 2.4.1 Investor Risks

The two nature transition risks of chief concern to financiers are a) reputation loss due to negative impact on the nature of their investments and b) unrealised financial gain from their investments. In this section we relate the impact that investments have on nature in the Amazon through to these transition risks, which are more salient to financial institutions.

**[a] Reputation Loss** - There are two ways within this model which we look at reputation loss. This could be from legal risks arising from non-compliance with the Forest Code. Alternatively, nature-related physical impacts across the farm generated by investment could pose a reputational risk to the investor if highly negative, and we assess this using the Nature Risk Profile (NRP). We acknowledge that this is a narrow and highly simplified view of reputation. However, we hope it provides the reader with a conceptual insight into how to link reputational risk to the value chain and AI, whilst keeping the modelling mathematics to a minimum.

**[b] Unrealised Financial Gain** - A reduced return of the investment in the beef supplier due to, for example, embargoing of farms who were found to be non-compliant with the Forest Code, or a reduction in production rates due to physical nature-related impacts building up.

Probabilistic models for both reputational and financial risks will be presented in Sections 2.4.4 Reputation Loss and 2.4.5 Financial Loss. These models will show how an agent-based probabilistic model can be used to both derive the reputational and financial risks as well as provide a means to notionally connect the organisational supply chain, investor dependencies and impacts.

In the remainder of this section we will show how AI can be used to calculate these transition risks and fill data gaps required to monitor Forest Code compliance and to calculate the NRP.

### 2.4.2 Data Gaps

We have identified three data challenges related to the Forest Code: errors in the CAR self-reported data, the mapping and validation of Permanent Preservation Areas (APPs) and the granularity of the Estimated STAR score.

There are problems monitoring Forest Code compliance at farm-level. For example, the accuracy of CAR data has been challenged as the self-declared data from landowners has to be validated by governmental agencies, which is proving to be a very slow process.[4] Work is underway to validate whether a CAR report is reliable or not. The probability that the CAR report is accurate can be determined with AI (de Melo Borges et al., 2021), using data from official validated reports. We note that a Validation Module is being developed by the Brazilian Forest Service (SFB), with support from the Federal University of Lavras (Chiavari & Leme Lopes, 2015). Although this work is able to determine whether a report is accurate or not, it does not provide a method to determine where in the report the inaccuracy arises. AI methods exist to identify relevant covariants (Marwala, n.d.) and these methods should be able to rank the significance of inaccuracies in the CAR reports.

The Forest Code requires that Permanent Preservation Areas (APPs) are reported in the CAR. This includes riparian areas, springs and water sources which can be detected via satellite imagery (WWF, 2016). Thus, remote sensing can be used to validate this part of the CAR report. New approaches in AI should be able to automate the segmentation of both riparian and water sources from satellite imagery. For example, Figure 5 shows the result of applying a UNet (Ronneberger et al., 2015) segmentation algorithm to aerial imagery of hedges, water sources, fields and anthropological structures in Oxfordshire. This technology is sufficiently accurate to determine the width of hedges and should be able to determine the extent and quality of riparian areas and reserve strips also. Furthermore, as noted in the figure, the technology can detect water sources. Similar quality maps should be constructable using high resolution satellite imagery, for example, Maxar 15cm optical imagery (Maxar, n.d.). We note that UNet has been proposed for the detection of deforestation in the Brazilian rainforest (Bragagnolo et al., 2021).

---

[4] CAR data is categorised differently in the governmental database based on the level of validation it has undergone. See: https://www.car.gov.br/#/consultar

We have found one data challenge related to the Nature Risk Profile. The Estimated STAR score is not well suited to comparing between farms at the level of CAR data, as it is at a granularity of 5km$^2$. This is because it is generated from the IUCN Red List data, and is not intended to be used to track changes at a site level, rather to indicate areas of biodiversity significance within a landscape. On the ground data collection can help to add a more granular and timely assessment of location significance and could be integrated into a Calibrated STAR score.[5] Crowdsourced or remote sensed collection can help. For example, AI research is underway to apply computer vision and bioacoustic sensing to identify individual birds or mammals of the same species from mobile phone acquired photographs or passive acoustic sensors (Ferreira et al., 2020; Linhart et al., 2022; Tuia et al., 2022).

We note that current research at Oxford University's Leverhulme Centre for Nature Recovery (LCNR) is aimed at collecting ecological data at scale. The LCNR is investigating biodiversity characteristics at a range of scales using AI from local to national levels (contact Steven Reece), including invertebrate monitoring using bioacoustic sensors (contact Yadvinder Malhi), ecosystem function monitoring using drone-based LIDAR (contact Jesus Aguirre) and soil health monitoring (contact Emily Warner). AI work with Tom Harwood (Environmental Change Institute) aims to determine the degree of forest degradation using AI anomaly detection techniques and measurements relative to a range of pre-determined healthy forest fragments.

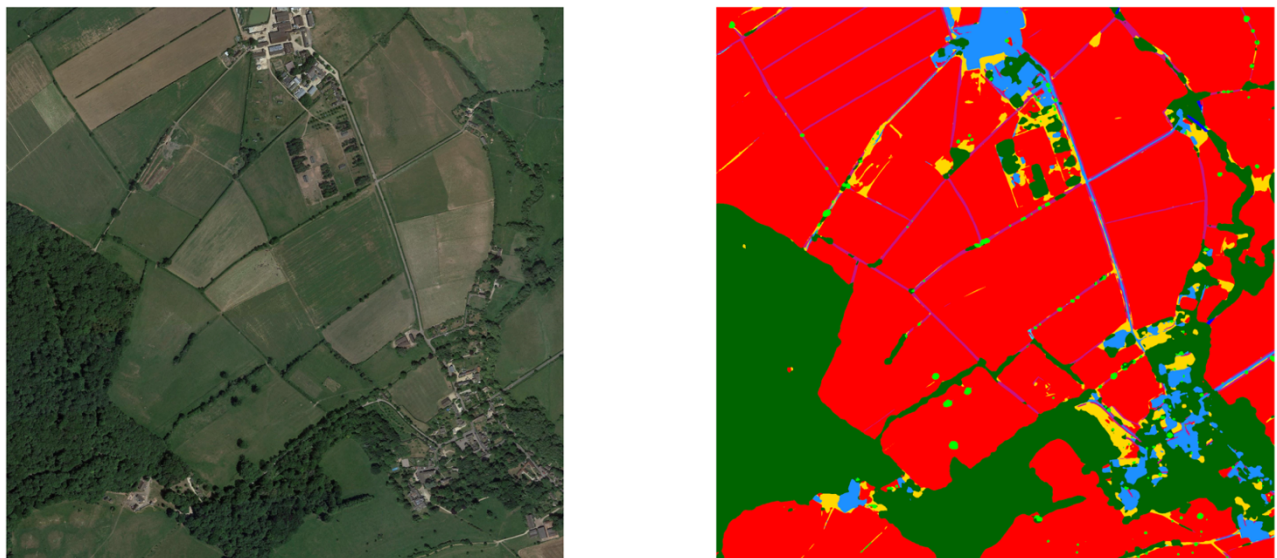

*Figure 5. Using UNet to map environmental assets: Left panel shows aerial imagery of Oxfordshire. Right panel is UNet segmentation of the aerial imagery into fields (red), water source (blue), wooded area and hedges (green) and man-made structures [Contact Steven Reece].*

---

[5] The Calibrated STAR score confirms the presence of threats and species at a site, establishing a baseline against which conservation management can be planned and targets set. For more see: https://www.ibat-alliance.org/star

So far, we have considered where AI can help improve the veracity of data for environment integrity monitoring. The impact investments have on nature can be inferred through Bayesian AI methods and the following subsections describe this proposed approach in some detail.

## 2.4.3 Modelling Investor Risks

We describe a model that could help financial organizations determine the impact they have on nature. The model determines which assets are likely to have high exposure to nature-related transition risks due to their impact on nature and their compliance with the local regulations. We use the model to simulate the value chain and determine environmental scores for each actor within this value chain. This allows the financial institution to decide whether to invest based on the nature-related risk in the value chain and to explore scenarios to manage the identified nature-related risk, for example through an SLL. The model is also useful because it encodes uncertainty and relationships between actors explicitly, providing transparent analysis.

We adopt a probabilistic framework to model the asset relationships within the supply chain. Probability theory allows us to express uncertainty in these relationships when not precisely known, to relate asset characteristics to observations or other indirect measurements of an asset, and to support supply chain and asset processes for time series analysis. We use $p(X)$ to mean our prior knowledge of $X$ and $p(Y \mid X)$ to mean $Y$ conditioned on $X$. We can interpret probability from a frequency perspective (i.e. how often $X$ occurs) or from a belief perspective (i.e. how much do we believe $X$). Both interpretations are valid.

The data described in the previous section is collectively represented as $D$, which comprises the raw satellite data described in Section 2.4.2 Data Gaps, derived products from Table 1 and any other data, including road network data, described later in this section.

The *state or condition* of the farm, $S$, is a vector containing the NRP impact score, data from the CAR report, and other derived features of the farm, such as the intactness of the Permanent Preservation Areas (APPs). Note that some of these products, including the riparian condition, are derived using AI from raw data, $D$, such as the satellite imagery. Any uncertainty in the AI derived product for farm $F$ is represented by $p(S \mid F, D)$.

Note, we can also include the social condition of the farm in $S$ if required. The social condition of a farm can include a measure of the adherence of the farm to workers' rights, or to its relationship with contested land. Farm social conditions are not considered further in this report. However, the probabilistic modelling paradigm presented in this report can be extended to accommodate social conditions.

As noted above, there are known issues with the accuracy of the CAR report. Furthermore, noise in the raw data or errors interpreting the data, or uncertainty arising from missing data (for example cloud cover in satellite imagery) lead to uncertainty in the state of the farm. Thus, for each farm we may only be able to specify a distribution over the possible states of that farm from the data: $p(S \mid F, D)$.

The exact form for $p(S \mid F, D)$ depends on the characteristics of the farm data and will not be explored further in this report. However, to illustrate how we can develop and refine this distribution, to include information about CAR accuracy, we will integrate the AI approach to detecting inaccurate CAR reports described in the previous section (de Melo Borges et al., 2021). The probability that the CAR report is accurate for a farm $F$, $p(CAR\ is\ True \mid F, D)$, can be determined using the AI method developed in (de Melo Borges et al., 2021), and data from official validated reports. To make use of this result we decompose $p(S \mid F, D)$ into two cases, where the CAR report is accurate, *CAR is True*, and where it is not, *CAR is False*. The distribution for the case where the CAR report is true, $p(S \mid CAR\ is\ True, F, D)$, can be determined using the AI methods described in the previous section. Thus:

$$p(S \mid F,D) = p(S \mid CAR \text{ is True},F,D) \, p(CAR \text{ is True} \mid F,D)$$
$$+ p(S \mid CAR \text{ is False},F,D) \, [1 - p(CAR \text{ is True} \mid F,D)]$$

From instances of farm for which their CAR reports are deemed inaccurate, and corresponding data of the state of these farms, it should be possible to determine p( S | F, CAR is False ) also. These factors can be integrated into the above equation to provide a distribution of the farm state.

### 2.4.4 Reputation Loss

There are two ways we look at reputation loss. Reputation loss can arise from legal risks for non-compliance with the Forest Code. Alternatively, nature-related impacts across the farm generated by investment could pose a reputational risk to the investor if highly negative, and we assess this using the NRP. We will look at both of these cases and formulate a probabilistic model that connects investments to their impacts on nature through the supply chain, that can be used to define reputation loss arising from the investment.

Suppose the investment is in beef supplier $B$. The investor's reputation should take a hit if their investment leads to a transgression of the Forest Code, or if the investment leads to nature impact with a low NRP impact score. We define the $L = 1$ if there is a Forest Code compliance, $L = 0$ for transgression and $N \in [0,1]$ for the NRP impact score. From the farm state we can determine if there has been a transgression of the Forest Code, and we can determine the NRP impact score with perhaps some uncertainty. We define $p(L = 1 \mid S)$ and $p(N \mid S)$ to be the probability that a farm in state $S$ complies with the Forest Code, or has an NRP impact score $N$, respectively.

We can now formulate a relationship between the investment and its impact on nature. For the Forest Code, loss of reputation is commensurate with the fraction of the investment that contributes to the farms that break the Code. Similarly, for the NRP impact reputation is commensurate with the expected impact the investment has on the greatest and most significant ecosystem. The probability that the investment in beef supplier $B$ complies with Forest Code, or has an NRP impact value $N$ are $p(L = 1 \mid B, D)$ and $p(N \mid B, D)$ respectively where:

$$p(L = 1 \mid B,D) \propto \sum_A \sum_S \sum_{F_D} \sum_F p(L \mid S) \, p(S \mid F,D) \, p(F \mid F_D,D) \, p(F_D \mid A,D) \, p(A \mid B,D)$$

$$p(N \mid B,D) \propto \sum_A \sum_S \sum_{F_D} \sum_F p(N \mid S) \, p(S \mid F,D) \, p(F \mid F_D,D) \, p(F_D \mid A,D) \, p(A \mid B,D)$$

and $p(A \mid B, D)$ is the probability the beef supplier sources from abattoir $A$, $p(F_D \mid A, D)$, the probability that the abattoir sources cattle directly from farm $F_D$ and $p(F \mid F_D, D)$, the probability that farm $F$ provides cattle through laundering (or other means) to farm $F_D$. Note that $p(F = F_D \mid F_D, D) = 1$ if farm $F_D$ does not source from any other farm. The probability that abattoir $A$ is a source for beef supplier, $B$, $p(A \mid B, D)$, can be obtained directly and empirically from Animal Transit Permits (Portuguese acronym GTA) (Moye, 2022). These permits are required by the Federal Government and show the movement of cattle from birth to slaughter. These permits also provide the probability that a farm is sourced *directly*, $p(F_D \mid A, D)$, but do not provide data on indirect cattle supply. Cattle laundering is modelled with the probability, $p(F \mid F_D, D)$, that cattle is transferred from farm $F_D$ to farm $F$ before being *indirectly* supplied to an abattoir. This probability can be determined using graph-based AI, the road network connecting farms to other farms and to abattoirs and openly available text reports published on Mighty Earth and Global Witness NGO websites (Tkachenko et al., 2023).

Note, in the equations above, we have assumed that the farm state is independent of the specific sourcing abattoir and that the probability of farm Forest Code compliance, or the NRP impact value $N$, depends only on the farm state and not, for example, on the sourcing beef supplier. The joint probability can be factored by assuming conditional independence. So, for the case of the Forest Code:

$$p(F, S, L \mid B, D) \propto p(A \mid B, D) \, p(F \mid A, B, D) \, p(S \mid F, A, B, D) \, p(L \mid S, F, A, B, D)$$
$$= p(A \mid B, D) \, p(F \mid A, D) \, p(S \mid F, D) \, p(L \mid S).$$

These assumptions should be verified by experts for each application of the model.

The advantage of *factoring* the joint probability can be seen from an expert modelling perspective. Each factor can be developed by the corresponding domain experts pretty much in isolation. For example, $p(L \mid S)$ is a legal matter, $p(S \mid F, D)$ an ecological matter, and $p(A \mid B, D)$ a supply-chain matter.

Note, although the model captures the proportion of investment that is compliant with the Forest Code, or has a high NRP impact score, it does not capture uncertainty in this proportion. A further level of probabilistic modelling is required for this.

## 2.4.5 Financial Loss

For simplicity's sake, in our model, the financial return depends directly on the number of cows processed by the supplier. This can be affected if a farm is embargoed, due to transgression of the Forest Code, for example. Here we define a measure of the financial return for the different situations of the supply chain, including possible farm closure.

Define $RET(B)$ to be the expected financial return to the investor by beef supplier $B$. This is the building block for alpha, the active return on an investment. The return could depend on the number of cattle, $C$, processed by each farm, the probability those farms are in the supply chain of abattoir $A$ and the probability that $A$ serves $B$. We will assume that the number of cattle processed depends on the state of the farm only. Note, it is possible to determine approximately the number of large mammals (including cattle) using AI and drone footage or satellite imagery (de Lima Weber et al., 2023; Duporge et al., 2021; Johnson, 2024; Laradji et al., n.d.). Furthermore, it may be possible to correlate cattle numbers with the size of the farm and the pasture quality, the latter determined using NDVI measurements.

$$RET(B) = K \sum_C \sum_A \sum_S \sum_{F_D} \sum_F C \, p(C \mid S) \, p(S \mid F, D) \, p(F \mid F_D, D) \, p(F_D \mid A, D) \, p(A \mid B, D)$$

where $K$ is a constant that captures the expected financial return to the investor per cow farmed.

The transitional risk models presented above are provisional and require detailed development and validation by experts before any further AI work is undertaken. However, we hope that these models provide some indication of how AI specifically, or modelling generally, can be used to relate the dependency and impact of the investment on nature.

## 2.4.6 E-Score

For each asset or organisation within the supply chain we can design an E-score, one from the Forest Code and another from the NRP, that reflects the impact the asset or organisation has on nature. Both E-scores reflect the current impact that the investment is having on nature. We assume that the E-score would be a value (or values) between 0 and 1 with 1 indicating a non-negative impact on nature and 0 a negative impact.

The investor E-score for the Forest Code would then be:

$$\text{E-score} = p(L = 1 \mid B, D)$$

and for the NRP:

$$\text{E-score} = \int_0^1 N\, p(\,N \mid B, D\,)\, dN$$

where $N$ is the Nature Risk Profile impact score.

Note, we can use the Bayes rule to derive the impact of all actors in the value chain. For example, substituting the marginals $p(\,L = 1 \mid A, B, D\,)$ or $p(\,N \mid A, B, D\,)$ into the above equations for the E-scores, we can derive the E-score for an abattoir given the investment. Alternatively, by marginalising all possible investment allocations, to yield $p(\,L = 1 \mid A, D\,)$ for example, we can derive the E-score for the abattoir given possible future investment allocations.

Because we can derive an E-score for each actor in the value chain, the model provides a transparent overview of the nature-related risk exposure attributed to each actor, rather than a composite overall score. This will allow financial institutions to better understand and manage their nature-related transition risk.

For future work we could extend the analytics to include positive impact on nature. (Within the scope of the current work we have focused on negative nature impacts, leading to nature risk. Neither the Forest Code nor the NRP currently track positive impact).

Note, the NRP E-score is aligned with TNFD's LEAP process. Identifying impacts and assessing their materiality are covered within the Evaluate stage, E2 and E4 respectively.

The following subsections provide a brief overview of what extensions are possible from the model.

### 2.4.7 Scaling the Model Regionally or Nationally

So far, we have modelled a single investor and the supply chain for abattoirs associated with the investor. The model can also capture all supply chains across all investors, beef-suppliers, abattoirs and farms both regionally and nationally. Thus, if an investor has a portfolio of investments over several beef-suppliers then the E-scores for the portfolio can be determined from the investment weighting for each beef supplier within the portfolio $p(B|\text{portfolio})$. Thus, for the portfolio the Forest Code and NRP E-scores become:

$$\text{E-score} = \sum_B p(\,L = 1 \mid B, D\,)\, p(\,B \mid \text{portfolio}\,)$$

and for the NRP:

$$\text{E-score} = \sum_B \int_0^1 N\, p(\,N \mid B, D\,)\, p(\,B \mid \text{portfolio}\,)\, dN.$$

### 2.4.8 From Asset Level to Organisation Level E-Scores

A consistent valuation of E-scores from data is important for their interpretation, temporal comparisons and use within financial decision making. So, there is value in transforming the asset level posterior E-scores derived using the Bayesian modelling methods described above to organisation level scores. However, the calculation of organisation level E-scores by ESG score providers can be obscure and the data used to obtain the scores can also be ill defined. Different ESG providers may obtain raw data from different sources, have different definitions of environmental indicators, or different ways of weighting these features. We hypothesise a graph regression algorithm (Tang et al., 2022) that learns the mapping from asset level scores to organisational level scores in a way that maintains consistent organisation level ESG score provider E-Scores. The algorithm would take the posterior Bayesian model as input (represented as a graph or network), along with the E-scores attached to each asset 'node' (Pearl, 2000). The output of the neural network graph would be a vector of E-scores for all the organisations in the supply-chain. The algorithm would be trained on organisation level scores provided by ESG providers and corresponding asset level Bayesian models.

### 2.4.9 Scenario Analysis with the Bayesian Model

The probabilistic model can be used to determine the impact of potential future government and investor decisions, amongst other considerations. We investigate two simple scenario cases here, and invite the reader to explore their own scenario requirements within their own models.

By extending our model to include a temporal dimension we can explore the consequences of strengthening legislation in the Forest Code, for example, leading to the embargoing of farms. An additional factor to the model, $p(F_t|L, F, D)$, would tell us the probability that the farm still exists at time $t$ given its state at a preceding time, conditioned on what we know about the farm (via data $D$) and legislation $L$. The stronger the law, the tighter this distribution would be around the embargoed farms (i.e. $p(F_t|L, F, D) = 0$) and consequently, we could determine the distribution of investor return at time $t$ under the new legislation.

For an investor to contemplate sustainable divestment in a particular beef supplier, the key consideration should be the impact of their individual investment on nature. However, the divestment hole could be filled by other investors. The impact on divestment can be determined by setting $p(A|B, D) = 0$ in the region level model discussed in Section 2.4.7 Scaling the Model Regionally or Nationally, and the model re-evaluated. It might be interesting to note here that embargoing farms, or extending the road network, would physically alter the supply network. Inferring the new supply network from scratch using the road network, abattoir and farm locations, as per (Tkachenko et al., 2023), should also include the financial and operational cost of switching suppliers.

### 2.4.10 Combining Corporate Controversy Reports

Appendix B provides examples of text-based reports pertaining to beef suppliers' compliance with the Forest Code or highlighting environmental misdemeanours. These reports can be used to reduce any remaining uncertainty in the reputational risk of an investment. Appendix B shows how these reports can be integrated into our Bayesian model.

### 2.4.11 Feedback Interviews with Relevant Organisations

We presented our model to organisations with an interest and expertise in some or all of finance, markets, laws and modelling their relationships, especially for environmental impact in Brazil. Of particular interest to us were related approaches already under investigation, validation of our approach and insights into possible extensions of our approach. The main findings are presented in this section. A summary of discussions with other organisations is presented in Appendix A.

*BVRio*
We presented to BVRio which is a non-profit with offices in Oxford and Rio de Janeiro. BVRio have developed the 'Soy Platform' which allows users to enter a specific CAR number and check if that CAR overlaps with deforestation events, protected areas, or indigenous or quilombo lands. Their tool is aimed at soy traders, as well as the financial institutions who fund them, to check if the soy they are buying is associated with illegal deforestation. Similar to our approach, the goal of BVRio's tool is to increase the transparency of the soy sector and the due diligence process for financial institutions. We discussed the potential of expanding the platform to beef in the Amazon and BVRio identified some current challenges they are facing which they think our model could help to address. Specifically, the BVRio team have found it difficult to track cows through the supply chain. They described that sometimes different farms do different parts of the process and most of the current checks only look at the last stage of the supply chain, so it is difficult to identify responsibility. Because our model looks at all actors throughout the supply chain, they thought that our model would help them tackle these challenges.

*Bank A*
We also presented our probabilistic approach to an international bank with activities in Brazil. It was interesting to note that they currently do not work at the asset level. They use IBAT to look up potential impact on biodiversity. Bank A was clear that a probabilistic approach to generating heatmaps of damage overlaid with where clients source from would be a good extension. This approach would help their risk teams, retailers, traders as well financial organisations. They also pointed out that our approach would be less useful for meat packers as they have their own proprietary monitoring systems and can monitor environmental compliance of farms in real time. Their issue is how to encourage the producers to be compliant. Bank A was keen to learn what our end product would look like, including the rate of heatmap update.

Bank A noted that Brazilian banks are often adept at modelling and monitoring environmental risk posed by individual farms. However, their limitation is in the indirect impacts through their supply chain for which they have little, if no knowledge. Furthermore, they stated that our approach would be useful for asset managers and organisations who are not based in Brazil.

We discussed E-scores with the bank, and they said they only use E-scores (derived by an external party) as a final check after an investment decision is made in principle, to ensure no major risks. What they found "really interesting" was being able to determine the nature impacts of a supply-chain which our model supports. They currently do not have this capability and are keen to know which trader is selecting farms to have a lower impact on the environment to work with. Making this information public and open would change the conversation they have with their clients. This was a valuable verification of our probabilistic modelling methodology.

*Quant Foundry*
We spoke with Chris Cormack, Managing Director of [Quant Foundry](Quant Foundry) and a Fellow of the [UK Centre for Greening Finance and Investment](UK Centre for Greening Finance and Investment). Quant Foundry provides model-based solutions for difficult to measure risks that combine quantitative modelling, data science and financial risk methodologies. They have experience in climate risk impact and ESG risk factors.

QF provided us with feedback on the viability of our modelling approach. They agreed that banks are trying to model regularity uncertainty and our approach is a viable way of embedding the transition risks. Furthermore, the approach allows policy makers to see if their decisions lead to something tangible, especially in Brazil, where biodiversity policy has already been enacted through the Forest Code. Governments could use the approach to understand where the risks are, and companies understand how they could be embedded. This would change the dynamic of risk management. However, our approach wouldn't apply everywhere, where such liabilities don't exist.

QF liked our agent-based approach which allows different model sophistication for each member in the value chain: whereas commercial businesses have detailed models, regulators are still getting to grips with AI and their models are simplistic. QF believes a key challenge problem is to build such models (or data selection) using AI that give the best investor outcome insight under uncertainty, especially uncertainty in the supply chains, by selecting an appropriate taxonomy of information.  This requires an economic framework which has yet to be determined.  QF is keen to collaborate with the research team on the probabilistic modelling and AI applications for policy and supply-chain networks.

*TNFD*

We spoke to James D'Ath and Laura Clavey from the TNFD Secretariat, who provided some feedback on the model. They mentioned that a large multinational bank operating in Brazil was looking for a way to trace cattle supply to the slaughterhouses they invest in. The banks don't have capacity to do this currently and consequently, although they could have environmental data it does not match up to their company's responsibilities in a way that they can make a strong financial decision. TNFD noted that our approach would address this problem ("it is the way to go"), and also help refine the analysis within their [LEAP framework](LEAP framework).

# 3. Use Case 2: UK Water Utility Sector

## 3.1 Introduction

The British water sector, responsible for freshwater supply and sewage treatment, underwent privatisation in the 1980s. Over the years, the sector has faced challenges in meeting the increasing demand from a growing population while dealing with ageing infrastructure, some of which dates back to the Victorian era (AMCS, 2024).

In recent years, individual water companies have been subject to significant regulatory fines, amounting to millions, due to mismanagement of sewage that resulted in the pollution of rivers and coastlines, leading to fish deaths and environmental damage (Morison et al., 2023). These high-profile incidents not only tarnish the sector's reputation in the eyes of investors and consumers but also impose substantial financial burdens, posing investment risks. To mitigate further financial losses and safeguard the sector's reputation, it is crucial to identify and monitor nature-related risks associated with water companies comprehensively. This enables investors to anticipate and manage risks while holding the companies accountable through investor stewardship and engagement.

Therefore, our case study on the British water utility sector focuses on developing predictive AI solutions to empower and enable effective nature-positive engagement by bondholders.

### 3.1.1 British Water Sector Context

Water companies in the UK are grappling with a range of risks (see Figure 6) and it's crucial for bondholders to take note of these factors to ensure effective management of environmental concerns in the medium to long term (see Figure 6). One key risk involves the growing pressure from consumers and policymakers for a nature-positive water supply chain. This push is prompting water companies to explore nature-based solutions to enhance their overall impact on biodiversity. To tackle these challenges, water companies are engaging with upstream farmers to implement nature-based solutions to decrease the nutrient load and pesticides entering their water treatment facilities. This approach not only lowers operational costs by reducing the need for chemicals required to process nitrogen and phosphorus but also brings wider benefits, including bolstering the company's reputation in an increasingly environmentally conscious landscape and reducing negative biodiversity impact.

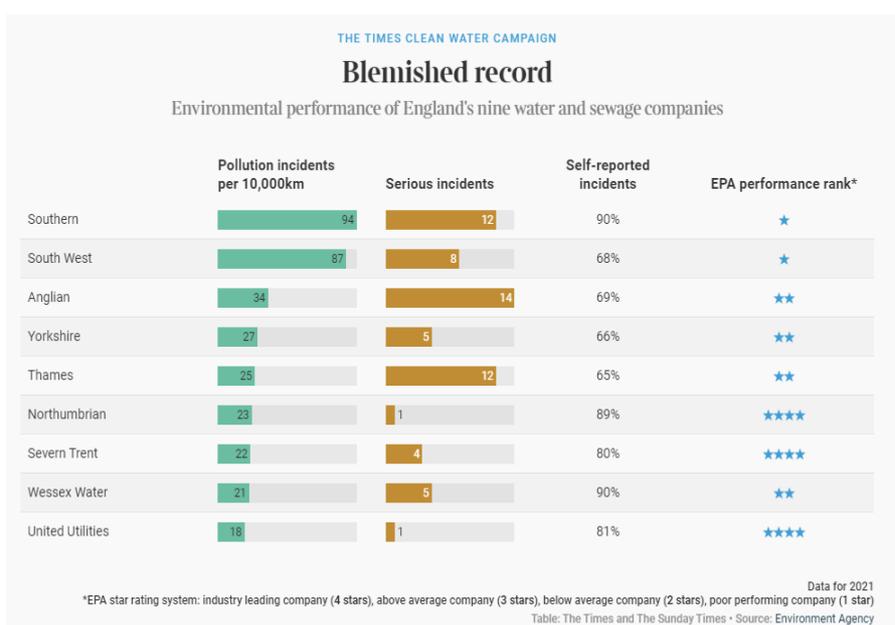

*Figure 6. The Times "Clean it Up" Campaign. The environmental record for UK water utilities in 2021 (Vaughan, 2023).*

### 3.1.2 Financial Context

The financial landscape for many UK water utilities has been turbulent since the 2020s. UK water companies have faced stark criticism based on their payout of large dividends to shareholders despite sustained sewage leakage and pollution issues as well as their complex and convoluted structures (Financial Times, 2023; Plimmer, 2024b). With the threat of larger fines from their regulator Ofwat, some water utilities are struggling to raise additional debt and equity to continue funding their operations and keep their business afloat amidst multi-million pound losses (Plimmer & Hollowood, 2023). There is not a clear consensus on the appetite for water utility-related investments by investors. The UK pension fund the Universities Superannuation Scheme decreased the value of its stake in water utility companies last year, which intensifies doubts in the asset's desirability (Plimmer, 2024a). However, FTSE 100 water group Pennon, which owns South West Water, bought Sutton and East Surrey Water in January 2024 for £89 million, which demonstrates continued appetite for the asset class (Plimmer, 2024b).

Traditionally, bondholders are considered to have less influence than shareholders in shaping corporate management and strategy. This is because bondholders lack the right to attend, ask questions, and vote in corporate annual general meetings, unlike shareholders. However, what often goes unnoticed is the leverage bondholders have over access to and the cost of capital, particularly during specific phases of the business cycle when corporations seek financing or refinancing (Caldecott, Clark, et al., 2022). Bondholders can negotiate investment terms and conditions, including setting nature-positive targets and influencing related strategies. As the water sector carries significant debt, bondholders play a vital role in shaping its direction.

### 3.1.3 Payment for Ecosystem Services

As part of their attempt to transition towards more environmentally responsible operations, water utility companies have introduced Payment for Ecosystem Services (PES) to incentivise nature-based solutions to manage water source quality. PES is a market-based mechanism to finance the protection of nature providing ecosystem services (IPBES secretariat, 2022). Traditionally, ecosystem services are provided by the environment for free as a public good and are an externality of the economic and financial system. In the PES model, the benefits that the ecosystem service provides can be internalised by paying the land-owner or manager to protect the natural resource and its associated ecosystem services (IPBES secretariat, 2022).

Water companies are implementing nature-based solutions to decrease the pollution upstream of their treatment facilities and reduce the nutrient load in the water requiring filtration (see Figure 7). Both the asset manager and the water companies are facing increasing pressure from policies such as the UK's biodiversity net gain target and the PR24 biodiversity baselining reporting, and from customers to improve their biodiversity footprint. However, water companies are struggling to convince the environmental regulator, the Environment Agency (EA), to permit them to employ untested nature-based solutions in particularly pollution sensitive areas. The proposed AI model could predict the efficacy of nature-based solutions by augmenting the spatially explicit planning capabilities of UKCEH's E-Planner and provide tangible evidence to the EA that nature-based solutions would be an effective alternative to engineered solutions.

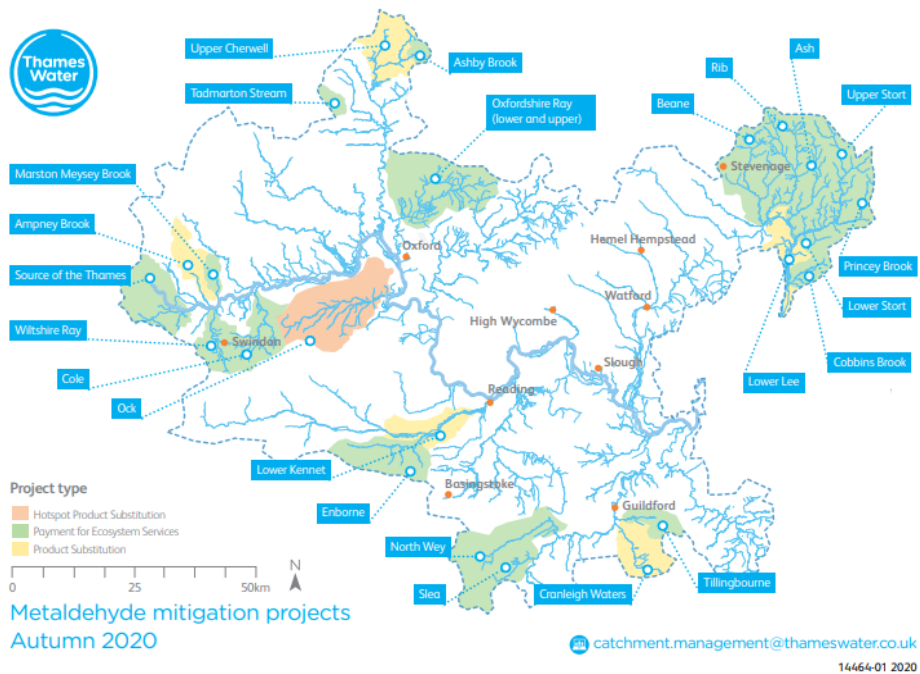

*Figure 7. An example of PES by Thames Water, the largest water utility company in the UK providing water to 10 million people across London and the Thames valley. Thames Water paid farmers to reduce their use of metaldehyde for slug prevention. In a results-based funding model, Thames Water paid enrolled farmers a premium if the water from their catchment concentration of metaldehyde was below a specified threshold (Thames Water, 2021).*

## 3.2 Methodology

There were two phases to this work: first research on defining the use case, through desk-based research and interviews with relevant financial institutions, and second, developing the AI model to fit this use case. We then refined the AI model by further conversations and research.

To explore the application of AI solutions to UK Water Utility Companies, we conducted interviews with a UK-based asset management firm that holds bonds issued by a UK water company to explore potential gaps in biodiversity data helpful for decision-making. After identifying potential gaps where AI could provide solutions, we also consulted ecologists at UKCEH and practitioners at ESG data providers to gain a greater understanding of the current data landscape. We also explored various other UK-based land use maps that use AI and satellite imagery to understand the work already under way in the area.

We consulted current grey literature to explore solutions put in place by UK water utilities, such as rules and regulations for nature-based solutions schemes for farmers. Desk-based research was also necessary to gain greater understanding of the UK water utility sector and its challenges in recent years.

Having come to grips with the context of the UK water utilities, the current biodiversity-related data availability, and the needs of asset managers for the sector, we began to construct an AI model. To do so, we consulted UKCEH on their current biodiversity-related data solutions and determined appropriate methodologies to bridge current gaps. Primarily, we want to allow for iterative decision-making and scenario exploration by the asset manager.

After completing the draft model, we presented the AI model to stakeholders for feedback. Having received constructive feedback from the asset manager on the usefulness of the model, we altered components to ensure the model was fit for purpose.

## 3.3 Results

### 3.3.1 Needs from Financial Institutions

We engaged with an asset manager (AM) that actively invests in debt instruments issued by UK water utility firms to discuss their current data usage and methods for assessing biodiversity risk.

The following points were raised in this discussion:
- The AM expressed concerns over the negative environmental impact of water utility firms and how that conflicts with the AM's own environmental, particularly biodiversity, targets.
    - Financially, the AM is concerned water utility firms are insufficiently managing their material environmental risks, resulting in costly fines.
    - Although as a debt holder the risk of non-payment is lower than shareholders, the AM nonetheless wishes to invest in financially sound companies.
        - *→ We propose developing a metric to effectively monitor the environmental impact water utilities to allow investment engagement to pressure water utilities to improve their biodiversity and ESG performance and risk by implementing nature-based solutions in their catchment areas.*

- The asset manager expressed frustration over the lack of reliable, externally verified, and audited data. Moreover, investors would like to have access to more temporally and spatially granular data.
    - *→ any metric developed should be spatially explicit and externally verifiable.*

- The asset manager also called for an investor guide with examples of data required, what investors can use to drive change, and guidance on interpreting biodiversity-related indicators.
    - *→ any metric developed should be intuitive to the financial community and easily interpreted and compared across investments.*

Some further relevant reflections are as follows:

- Framing nature losses and gains around the TNFD is helpful as investors and investees (in this case, water companies) can speak the same language and share common expectation of disclosure style and quality.

- It would also be helpful to produce an investor guide for biodiversity, similar to that for the Just Transition for Social Housing Companies, a document for the wider market on examples of what is needed, and what investors can use to drive change.

- It would be possible to produce a report for water companies for biodiversity in the same way they did for Network Rail.

- Water companies have a lot of conflicting priorities and are keen on Sites of Specific Scientific Interest (SSSIs) and scoring, and biodiversity unit scores (simple metrics) consistent, robust and simple and provide direction of travel. A practical approach might therefore be to calculate the baseline biodiversity unit score using the Defra Metric 4.0 methodology for the land that the water company either owns or has under management agreement. This could form the basis of simple metrics and targets. For example, no net loss of biodiversity (as defined by no change in biodiversity unit scores for X proportion of catchments) by 2025 moving to 5% net gain for these catchments by 2030.

### 3.3.2 Examples of Current Data Usage

Although there is reporting on environmental and financial performance of UK water utilities from internal disclosures and external regulators, quantitative, auditable and real-time data for monitoring is missing. As biodiversity is inherently local, spatial data is incredibly important in enabling financial institutions to effectively mitigate and manage biodiversity-related risks and opportunities.

Furthermore, real time data is integral to allow engagement on the most up-to-date issues and propose forward-looking solutions. Without much needed quantitative monitoring information, bondholders, such as asset managers, cannot effectively engage with UK water utilities because they do not have the ability to measure progress against biodiversity-related targets or determine success of implemented solutions.

In Table 2 below we detail the existing data sources that the asset manager use for assessing water utilities on their environmental and biodiversity performance. We have also identified other data sources that might be useful to fill in identified data gaps, including temporal and spatial granularity for monitoring.

*Table 2. Survey of relevant environmental datasets and methodologies*

| Data Source | Description | Project Relevance |
|---|---|---|
| [Water Situation Score](#) | An example of a score produced by an asset manager to assess the ESG performance of water utilities for appropriate engagement as a bondholder. The asset manager scores the ESG performance of water utilities against 17 proprietary indicators, which include climate adaptation and biodiversity risk (Gibbs & Chiu, 2023). | The Water Situation score demonstrates the current approach by asset managers at measuring biodiversity risk, which is a component of their larger ESG risk assessment. *Issue*: The Water Situation score is currently based on backwards looking data and annual reporting, such as the Environment Performance Assessment of water utility regulator Ofwat (Gibbs & Chiu, 2023). |
| **Water Utility Corporate Sustainability Disclosures** | The UK water utility companies produce voluntary annual ESG and sustainability reports with information about their progress and performance on ESG-related targets and metrics. | These sustainability reports are examples of the kind of data that asset managers rely on to provide utility specific information on their ESG actions and performance. *Issue*: There is limited spatially explicit data provided, limiting the usefulness for meaningful biodiversity risk assessment. Additionally, these reports are voluntary and will often not provide quantitative or auditable metrics, making it challenging investors to accurately assess biodiversity risk or action. |

| [Environment Agency's Annual Environmental Assessment](#) and [Ofwat's Annual Performance Report](#) | Annual reports on the performance of UK water utilities by their regulators. These reports include information on their environmental performance against indicators, such as pollution incidents, sewage treatment, and future improvement plans, among others. The reports are issued once a year for the previous year. | Another example of data available for investors to assess the ESG or biodiversity risk of UK water utilities. **Issue:** Again, there is limited spatially explicit information provided. Also, the reports reflect performance the previous year with no real-time or forward-looking data for investors to assessment current or future ESG or biodiversity risk exposure. |
|---|---|---|
| [UKCEH E-Planner](#) | E-Planner is a spatial decision support tool to support environmental decision making at the field, farm and landscape scales. The tool combines high resolution (5m) environmental data describing soils, topography and the location of high nature value habitats in the landscape to calculate relative suitability scores for the following habitats: flower-rich habitats for pollinators; wild bird food patches; water resource protection; natural flood management and afforestation (Redhead et al., 2022). | This tool can play an important role in implementing nature-based solutions in water catchments to improve water utility ESG and biodiversity-related performance. **Issue:** The tool currently does not have scenario capabilities and does not have forward looking or predictive components to assess different courses of action for engagement. |

### 3.3.3 Key Priorities Identified by Asset Manager

*Ideal data availability*

- The asset manager is primarily interested in obtaining independent, water discharge and water quality data.

- Additionally, the asset manager is interested in attributing chemical pollution to the asset level (e.g. at the farm level).

- Need a natural capital baseline from which to measure change. There was one example of a water company doing this but it is not common at the moment.

- Simple, consistent metrics aligned with key performance indicators would be useful to measure change.

*Solutions*

- Various solutions were suggested. For example, water quality monitoring stations can be set up to obtain real-time measurements. Spot checks can also be carried out at critical juncture points of a waterway. However, there are costs associated with such methods.

- Nature-based solutions were suggested as a low-cost and longer-term solution. For example, monitoring the added bend to rivers will act as an indicator of effective waste control. The asset manager adds that such solutions will earn carbon credits and biodiversity net gain credits for water companies.

- Furthermore, the asset manager reacted positively to additional land use, agriculture, species, and habitats data provided by CEH Wallingford. At this stage, the asset manager is interested in showcasing these datasets to their investees, by way of pushing for more comprehensive, and externally verified data source for reporting. This includes the monitoring of Sites of Special Scientific Interest (SSSIs). We note that SSSIs are monitored by Natural England. However, water companies could do this themselves using UKCEH E-Surveyor (UK Centre for Ecology & Hydrology, n.d.-b), for example, to record indicator species for the SSSIs.

## 3.4 Proposed Model

### 3.4.1 Investor Risks

The two transition risks of chief concern to financiers are a) reputation loss due to fines arising from environmental pollution b) significant costs arising from fines, chemical clean-up and NbS implementations. In this section we relate the impact that investments have on nature in the UK through to these transition risks, which are more salient to financial institutions.

**[a] Reputation Loss** - Environment-related fines and penalties can be harmful to the reputation of a company. It is challenging to directly, quantitatively attribute the financial repercussions of reputational risks, but there are several avenues whereby poor reputation can lead to financial losses and undermine the long-term survival of the company. In the UK, individual households cannot independently choose their water and sewage service provider, which takes away the risk of losing customers. However, poor reputation may place water utility companies under greater scrutiny by the media and attract further criticism and lobbying from civil society organisations. Second, as the result of long-term, repeated environmental breaches, a company may have to go through governance and leadership changes, which can undermine the stability of corporate operations, and ultimately bear negative repercussions on the prospects of it being able to provide satisfactory water treatment services. In our model, we assume that the more frequently a company is fined and the larger the fine, the greater reputation loss the company faces.

**[b] Indirect financial losses** - Bondholders are unlikely to be directly affected by the fines and penalties faced by water utility companies as the result of environmental pollution. This is because the financial cost of the fines, though significant, are unlikely to lead to debt default (meaning the bondholder will not be repaid the principal and interest in full). However, in recent years, given the emergence of active credit investment strategies (trading rather than buy and hold), bondholders increasingly view their investments as volatile assets whose value fluctuates based on performance (Schwarcz, 2017). Fines can erode the profitability of a business, and the financial performance of the issuer becomes increasingly important as it signifies the ability of a company to invest in new technologies, better services, research & development, all of which can contribute not only to the long-term growth of the company but also their environmental performance overtime. In other words, any negative impact incurred on a water utility company's balance sheet as the result of fines is going to be material for bondholders.

Probabilistic models for both reputational and financial risks will be presented in Sections 3.4.7 Reputation Loss and 3.4.8 Financial Loss.

Our proposed model represents an optimization problem to strike a balance between the expense of fines and operating costs compared to payment for ecosystem services. This approach allows us to include a time series approach to provide opportunities to change strategy at given intervals, matching the intervals of bond (and loan) refinancing. This proposed model was conceptually developed but not tested or validated, which was beyond the scope of the current project.

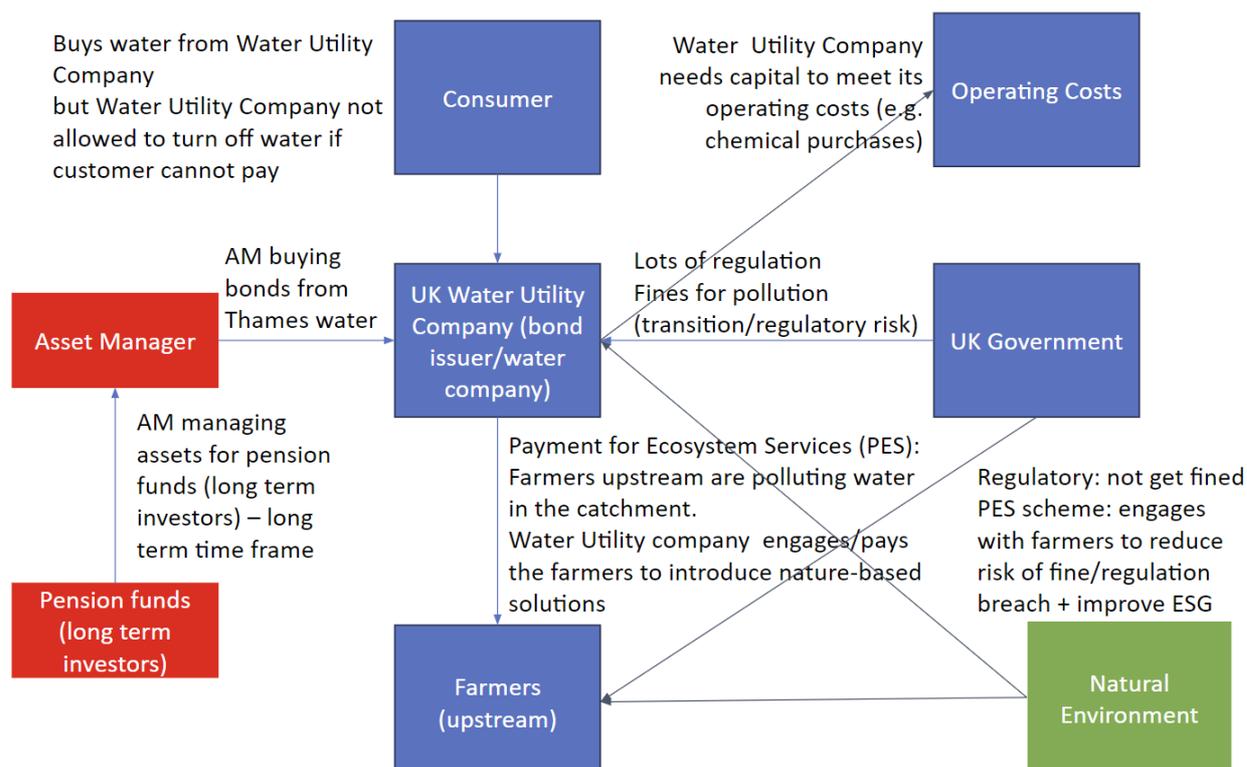

*Figure 8. Relationship diagram for actors in the water utility company NbS bond scheme.*

In the chosen use case, pension funds have paid an asset-manager to invest their capital to receive the appropriate returns on the appropriate time scale, typically looking for low risk and long-term investments (see Figure 8). The asset manager buys bonds in a privatised UK water utility company. The UK government regulates water utility companies' performances with targets for pollution and leakage, among others, and levies fines against the utilities if those targets are not met. This is a form of nature-related transition risk as the Government has set policy to limit the damage to the natural environment. Consumers are affected by performance failures because they pay water utility companies for clean and reliable water and will be impacted by any price rises due to financial instability or insolvency. Both water utility companies and farms impact the natural environment by releasing pollution into the surface water and groundwater, degrading the local ecosystems. Water utility companies require adequate capital to cover their operating costs, such as purchasing purifying chemicals, running their waste treatment plants, necessary infrastructure upgrades and maintenance. In this use case, water utility companies pay upstream farmers to implement nature-based solutions (NbS) to reduce their water pollution, which can reduce the load on the water treatment facilities and decrease the operating costs of water utility companies. The asset manager, who invests in water utility bonds, hoping to protect their investment and respond to increasing customer pressure for biodiversity positive investments, is increasing pressure on water utility companies to reduce their pollution and negative biodiversity impact through a PES scheme.

We examine two potential NbS options from a water utility company's corporate report, the Catchment Fund Handbook, their requirements for implementation, disclosure and milestone monitoring (Thames Water, n.d.). Both NbS require farmers creating and maintaining buffer strips in order to reduce the run-off of nitrate and/or pesticides into neighbouring water-sources.  Buffer strips intercept surface runoff, reducing the risk of contaminants reaching watercourses. Removing land from cultivation means no fertiliser is applied on the buffer, reducing the risk of leaching of nutrients to groundwater.  Buffer strips can also enhance biodiversity, providing valuable habitats for flora and fauna.

In both NbS cases, E-Planner, developed by UKCEH, provides a score between 0 and 1 for four environmental variables in their *Opportunities maps* (see Figure 9) which, collectively, provide the potential applicability of the NbS.  The indicators include topography, existing habitat features, environmental assets and soil erodibility.  Two immediate AI-driven extensions to E-Planner have been discussed with CEH Wallingford:

- AI to combine maps of crop planting history with the E-Planner maps to predict water quality arising from rainfall runoff from fields.  This data-driven approach would require water quality measurements immediately adjacent to the field and some knowledge of NbS existing interventions in the field.

- Adding a layer to E-Planner that uses AI to predict the spatial appropriateness of an NbS.  This would be a classifier trained using E-Planner maps and farmer supplied NbS selections from these maps.

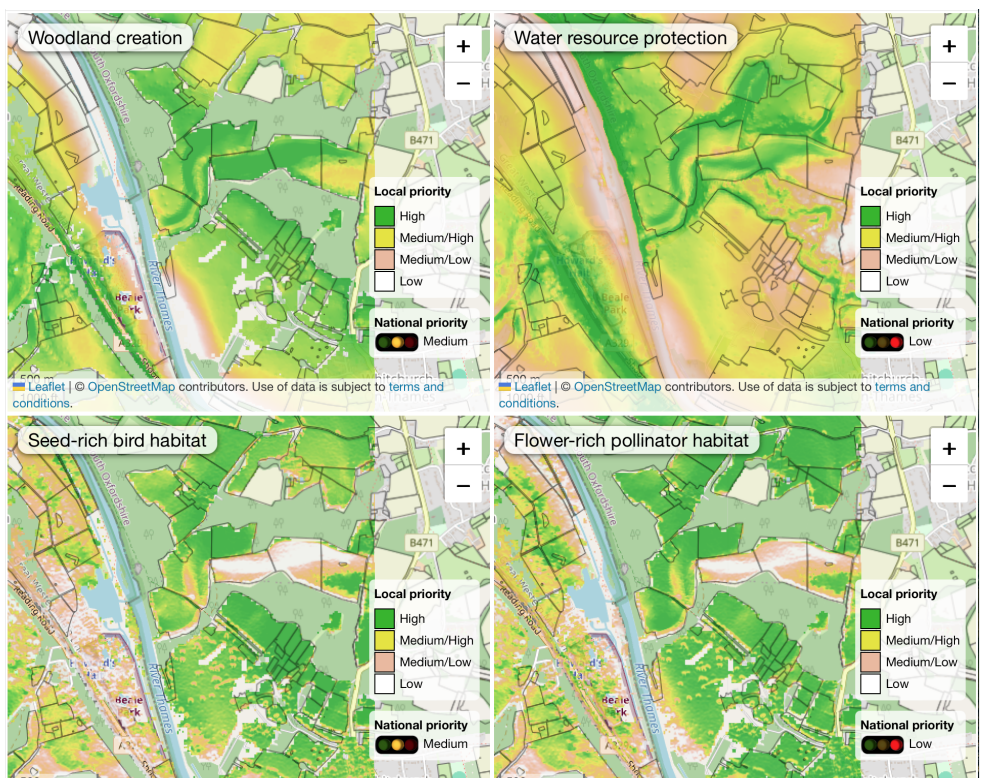

Figure 9. UKCEH E-Planner Opportunities Maps

4m-6m buffer on cultivated land
This NbS option is for establishing areas of uncultivated land between the productive part of the field and an existing feature or habitat like a hedgerow or watercourse to improve the water quality by reducing pesticide and nitrate runoff.  This option applies to land managed in rotation including temporary grass.  Permanent pasture is not eligible.  Farmers can apply for funding for new buffer strips or for maintaining existing buffer strips unless they are already in an existing scheme and already being funded. UKCEH provides 10m resolution land cover maps, derived from Sentinel-2 satellite imagery using an AI random forest algorithm (Marston et al., 2022). These maps (see Figure 10) can be used to determine the existence of temporary grass and therefore can validate the land is used in rotation (Upcott et al., 2023).

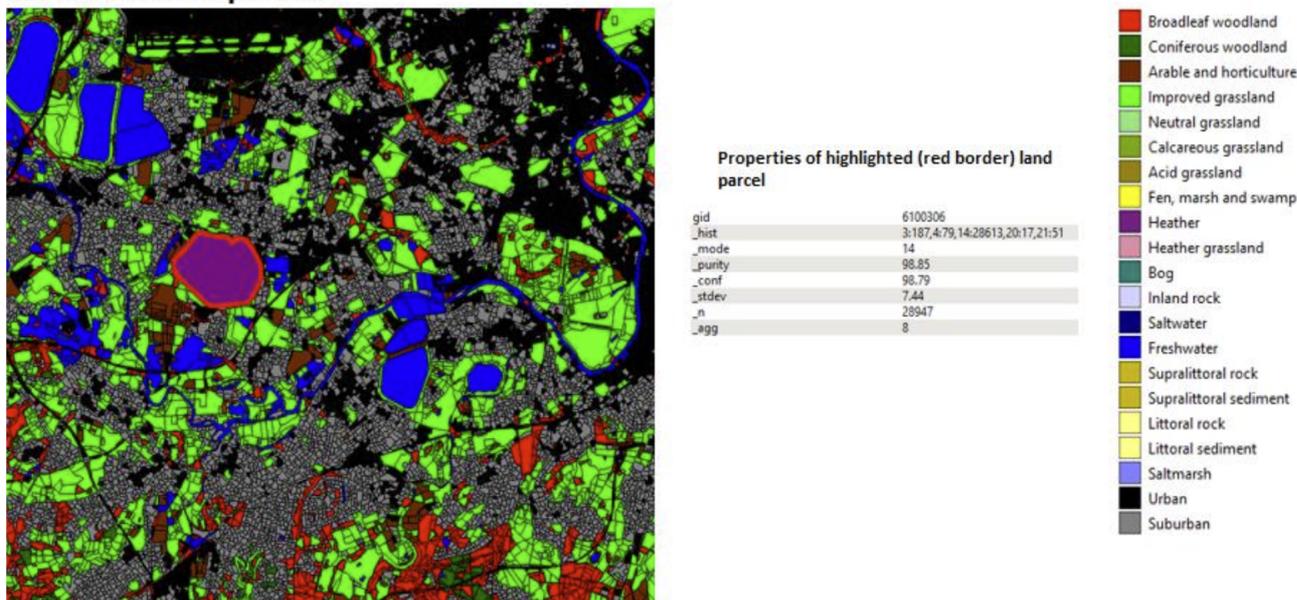

*Figure 10. UKCEH Land cover map: classified land parcels and properties of the highlighted (red border) land parcel.*

The general requirements to implement the NbS are provided next and we provide some indication of the data source and AI capability to monitor these requirements. We note that the farmer is legally required to provide supporting data throughout the NbS implementation and this data, chiefly photographs, can be analysed via AI. Farmers must supply:

- Maps showing where they intend to establish the buffer strips and any local watercourses or ditches. It is possible to verify existence of watercourses from high resolution satellite imagery,

- Photographs of the fields where they plan to establish the buffer strips,

- Details of the seed mix they plan to sow (if applicable),

- With milestone forms the farmer must provide:
    - Proof of purchase for seed mixes sown (if applicable),
    - Field records including, but not limited to, areas and dates sown, how the option fits in with rotational cropping, records of any applications made, herbicides applied, method and dates,
    - Photographs of areas prior to sowing, during and once fully established.

The general requirements for implementing this NbS are:
- Sow a grass/wildflower mix with a minimum of six grass/wildflower varieties. *We note that, for NbS funding, seeds currently cost £180.50 per hectare. It may be possible to adapted Region-based Neural Network approaches (e.g. Faster-RCNN) (Abbas et al., 2022) to locate and classify wildflowers in photographs submitted by the farmers as part of their milestone reporting. However, current deep learning approaches work for up close photographs of the flowers. Note, this approach is an extension of the technology* *used in UKCEH's E-Surveyor which deploys a convolutional neural network to classify images of individual flowers.*

- Cut 1m-3m on the edge nearest the crop each year after 15 July. *Again, this may be monitored automatically using AI through the farmer submitted photographs by segmenting the buffer regions using UNet. Furthermore, monitoring can be complimented via satellite imagery (see Ecological Engineering 94 (2016) 493–502 for a very early application of neural networks in this field. More advanced deep segmentation algorithms, such as the UNet, should offer more accurate results). Finally, OS Master Maps can be used to verify that pylons and other disruptive infrastructure are not present in the potential cut.*

- Buffers must be in arable fields and clearly marked on the submitted maps. *Arable fields can be distinguished from pastureland using UKCEH Living England National Scale Habitat Map (Kilcoyne et al., 2022).*

- Farmers are not permitted to:
    - Cultivate the buffers once established. *Cultivation can be determined from satellite derived NDVI time series as only uncultivated land is green all year. Cultivated land will show a clear difference in NDVI between sowing and harvesting (Alison Smith, personal communication. See also* (Klimavičius et al., 2023) *for example). These maps can be derived from Sentinel-2 imagery.*[6]

    - Use fertilisers, manures, insecticides, fungicides, or lime on the grass buffer. Herbicides may only be used to control weeds or to spot treat injurious weeds, invasive non-native species, nettles, or bracken. Blanket spraying is not permitted, only targeted treatment of affected areas. *Farmers regularly collect precision agriculture data including GPS coordinates of their tractor fertilisation activities. This information, if made openly available, can be used to monitor the spraying extent. Furthermore, written records of spraying often produced by farmers, if made available, could be used to monitor the extent of spraying.*

    - Graze the buffers. *No source identified.*

    - Use the buffers as access ways for livestock or machinery. *No source identified.*

    - Cut the buffer, except to control woody growth and for edge maintenance. *No source identified.*

    - Remove fallen timber or the limbs of any hedgerow trees over 30cm in diameter. *The absence of a tree can be gleaned from a sequence of the UNet derived maps from aerial, high-resolution satellite, imagery or possibly the farmer supplied photographs. Once the disappearance of a tree is noted then human intervention can determine whether the timber is still in place.*

    - Use this option where it might overlap a public right of way. *The locations of rights of way in the UK can be obtained from* [FootPathMap](FootPathMap).

    - Relocate buffers during the agreement. *Again, the location of buffers can be determined from NDVI maps obtained from Sentinel-2 satellite imagery.*

Some of the criteria above are difficult to monitor individually. However, the overall farmer requirement is to maintain healthy buffers to mitigate toxin runoff. Changes in NDVI will pick up these disturbances. The only natural disturbances, caused by seasonal flooding and water cover, can be detected using satellite imagery.

---

[6] *https://www.arcgis.com/home/item.html?id=dccafe125bbe4e2bb3315393acbd4701*.

4m-6m buffer strip on improved permanent grassland
This option is for establishing areas of uncultivated land between the improved permanent grassland and an existing feature or habitat like a hedgerow or watercourse to reduce nitrate runoff. This option can be used adjacent to intensively managed, improved grassland fields receiving more than 120kg/ha of nitrogen per year from fertilisers or manures. The farmer can apply for funding for new buffer strips or for maintaining existing buffer strips unless they are already in an existing scheme and are already being funded. They cannot use this option where it might overlap a public right of way.

Buffers should be established after harvest and after the bird breeding season (1 March to 31 August). This option requires a subset of the general requirements and supporting information for the cultivated land option above. Thus, the data requirements are identical to those in the previous option except:

- The farmer must leave an uncut buffer strip 4m-6m wide around the edge of any fields that will be mown (livestock can graze the uncut strip). *Again, this can be monitored automatically using AI through the farmer submitted photographs or satellite imagery.*
- If required, cut 1m-3m on the edge nearest the improved grassland each year after 15 July. *Optional so not monitored.*
- Only cut the sward in the buffer to control woody growth. *No source identified.*

The supporting information required from farmers for this option is very similar to the buffer on cultivated land NbS option above. The only significant addition is a map showing where the farmer intends to establish the buffer strips and any local watercourses or ditches. The presence of water features can be obtained from UKCEH land cover map.

### 3.4.2 Modelling Investor Payment Schedule

Our model is aimed at helping water utility companies to invest in NbS through bond issuances across all farms in the water utility company's catchment area. This involves selecting the most cost-effective NbS for each farm (or field), that sufficiently reduces the long-term nitrate and pesticide rainfall runoff from cultivated and non-cultivated land. The farms are monitored, as per the requirements above, as the NbS are established. At predetermined time instances, the water utility company may decide to change its NbS funding strategy if, for example, farmers have not satisfied the NbS funding requirements. Thus, the company may switch to a different optimal NbS strategy at key points, such as renewal of NbS contracts with farmers (annually) or the termination and potential re-issue of the bond. At these key points, individual farms may switch, or lose funding for, their NbS over time. The time series nature of our probabilistic model is a key difference to the Brazil Cattle Ranching use case.

Similar to the Brazilian Cattle Ranching Use Case, the value of our model lies in its ability to attribute impact to each catchment in the value chain. This allows for the financial institutions or corporates to clearly identify areas where nature-based solutions are consistently delivering positive outcomes for nature and to allocate capital efficiently to successful nature-based solutions in these areas at each time-interval.

### 3.4.3 Estimating River Pollution Condition

In this section we propose a model to predict the pollution from each field over each time interval t, as a function of the farm condition, the NbS and rainfall experienced over the time interval.

We define $N_f$ to be the sequence of nature-based solutions implemented in field f. The NbS variable $N_f$ is a detailed description of the nature-based solution and could include the toxin absorption rate of the NbS at different stages of its development. The field state $S_f$ includes information on the extent of the toxicity of the field (phosphorus and nitrogen). Over time interval $t$, for which there was total rainfall $Ra_t$, the pollution $\Pi_{f,t}$ from the field is a function $g$:

$$\log \Pi_{f,t}(N_f) = g(S_f, N_f, Ra_t).$$

Note that $\Pi$ aligns with the TNFD "Volume and concentrations of key pollutants in the wastewater discharged" metric for the agriculture and food sector (TNFD, 2023c).

The function $g$ is a geophysical model and can be physics-based, a combination of meteorological and hydrological models, when known, or it may be inferred from data, for example, by assuming that $g$ is a Gaussian process (Rasmussen & Williams, 2006), or $g$ can be a hybrid data-driven and physics-based model (see (Reece et al., 2014) for an example of a hybrid physics-based/Gaussian process model). We will not consider the details of the geophysical model here but note, as mentioned in Section 3.4.1 Investor Risks, we have discussed such a model with UKCEH as a possible extension to their E-Planner tool (UK Centre for Ecology & Hydrology, n.d.-a).

The total pollution from the catchment into the river during time interval $t$ is from all fields in the catchment:

$$\Pi_t(\{N_f\}) = \sum_f \Pi_{f,t}(N_f)$$

where $\{N_f\}$ is the set of NbS for all fields in the catchment. Note, if rainfall can be forecast (seasonal variation, for example), then we can use $g$ to predict the toxin runoff with corresponding uncertainty from future farm state, rainfall and future NbS condition. This information would be required to allocate funds to the farmers.

### 3.4.4 Chemical Clean-up Costing

Over the time interval $t$ the cost of chemicals to clean the river water is proportional to the amount of pollution, $\Pi_t$, in the river $\rho(\Pi_t)$.

### 3.4.5 Regulation Fines for Pollution

During the time interval $t$ the government regulatory fine $F_t$ is proportional to the river pollution over the previous time interval $t-1$. However, the fine is set so that the water company can maintain it's operational capability. We assume therefore, that the fine is a function of the company's previously reported balance sheet as well as the river pollution, $F_t(\Pi_{t-1}, B_{t-1})$, where $B_{t-1}$ is the company's balance sheet. The balance sheet time series can be estimated using the recursive formula developed in 3.4.8 Financial Loss.

### 3.4.6 NbS Payments

The NbS payment over time interval $t$ for NbS $N_f$ in field $f$ is $pay_{f,t}(N_f)$. The scheme cost over time interval $t$ over all fields in the catchment is therefore:

$$C_t(\{N_f\}) = \sum_f pay_{f,t}(N_f).$$

### 3.4.7 Reputation Loss

The reputation loss is a function of the cumulative fines that the government has imposed on the water utility company due to poor river water quality. The reputation at the end of time interval $t$ is a measure between 0 and 1 with 1 indicating excellent reputation:

$$Reputation(t) = exp\left(-\sum_{i=1}^{t} F_i(\Pi_{i-1}, B_{i-1})\right).$$

Note, this is the reputation for a single instance of $\Pi$ and $B$. Of course, there is uncertainty in all of the model variables: pollution $\Pi$ is a draw from a Gaussian process and government-imposed fines and even the NbS costs may also be uncertain. Given the potential non-conjugacy of the corresponding probability distributions, and the potentially non-linear nature of the functions $F$ and $\rho$, a Markov Chain Monte Carlo (MCMC) approach to inference and estimation would be the first port of call (Bishop, 2006). Through MCMC the reputational risk, that is the expected value of the reputation loss, can be calculated. Further validation of the model, and details of the functions in the model, are required before this approach is developed further.

### 3.4.8 Financial Loss

The water utility company's balance sheet is positively impacted by the income received by providing domestic clean water, but negatively affected by fines, chemical clean-up costs and funding the NbS initiatives. Of course, the water utility company wants to maximise its balance sheet and, as noted above, can do so by carefully selecting which NbS to deploy.

The company's balance sheet $B_t$ at the end of time interval $t$:

$$B_0(\{N_f\}) = initial\ balance,$$
$$B_t(\{N_f\}) = B_{t-1}(\{N_f\}) + I_t - C_t(\{N_f\}) - \rho\left(\Pi_t(\{N_f\})\right) - R_t - F_t\left(\Pi_{t-1}(\{N_f\}), B_{t-1}(\{N_f\})\right)$$
$$- other\ expenses$$

where $I_t$ is the water utility company's income over time-period $t$ and $R_t$ is the bond repayment. Note, following our conversation with Insight Investment (see the next section), in some situations the bond repayment interest would be tied to the efficacy of the NbS and the river water quality. Thus, $R_t$ would be a function of the river pollution $\Pi_t$.

Since the company wants to maintain a positive balance sheet at each time $t$ then the problem is to find the allocation of NbS for all fields (note some fields/farms may not develop an NbS) that maximises the balance sheet subject to $B_t \geq 0$:

$$argmax_{\{N_f\}} \sum_t B_t(\{N_f\})\ such\ that\ B_t \geq 0\ for\ all\ t.$$

Again, when the model parameters are uncertain we can use MCMC to find the distribution of balance sheet optima.

### 3.4.9 E-Score

An E-score can be designed to reflect the current or planned future condition of the river water and farming environment. Although chemical cleaning and the NbS should achieve similar outcomes on the cleanliness of the river, the NbS should be preferred from an environmental perspective as it also provides a natural habitat. Thus, we propose an E-score that captures both the cleanliness, $\sigma(-\Pi_t)$ [where $\sigma$ is the sigmoid function], of the river over the interval $t$ and the proportion of toxin mitigation due to the NbS. Note, $\sigma(-\Pi_t)$ is bounded between 0 and 1 and could be predicted or directly measured as per Section 3.4.3 Estimating River Pollution Condition. So, the E-score at time $t$ is:

$$\text{E-score}(t) = \frac{C_t(\{N_f\})}{\rho(\Pi_t) + C_t(\{N_f\})} \times \sigma(-\Pi_t).$$

The E-score is 1 when the river is clear of pollution and no chemical water treatment is undertaken, it is 0 when either the river is highly polluted or no NbS is used to mitigate rainfall runoff pollution and is between 0 and 1 for other cases. Note, the expected E-score (calculated using MCMC, for example) is an alternative score when the model variables are uncertain.

The E-score evaluates the efficacy of the investment in nature-based solutions in each catchment. This allows greater granularity for the water utility company to make decisions about where to invest in nature-based solutions and providers greater certainty to the financial institution for informed engagement on nature positive outcomes.

### 3.4.10 Scoping Interviews with Relevant Organisations

We presented our model to organisations with an interest and expertise in some or all of finance, markets, laws and modelling their relationships. Of particular interest to us were related approaches already under investigation, validation of our approach and insights into possible extensions of our approach. The main findings are presented in this section. A summary of further discussions with other organisations is presented in Appendix A.

*Asset Management Firm*
An asset management firm that regularly invests in water utility companies noted that storm overflow damage is more of a concern than rainfall runoff from farms and, although they were uncertain about the Environment Agency's policy on fines for phosphorus and nitrogen pollution, reduction in these toxins would reduce the costs for the water utility company down the line to treat the water. This led us to refine our model and introduce water chemical costs (and also the potential for other water improvement costs such as ceramic water filters). However, representatives from the asset management firm stated that if the chemical clean-up costs were less than the NbS costs then the NbS won't be implemented in real life unless there was customer pressure or the water utility company had a desire to be a better class of water company. Furthermore, if the water utility company's operation is connected to water quality permits then an NbS would not be operationally possible. Finally, representatives from the asset management firm introduced us to 'PR24', where the water utility company submit their business plan for the next 5 years to Ofwat. Part of this is a detailed plan for capital expenditure and they go through a very rigorous process where they would need to argue for spend on biodiversity positive initiatives. Many water companies already have detailed models for this, with in-house hydrologists but their models do not translate to financial risks. So our project is very innovative.

*Environmental Change Institute (Oxford University)*
Alison Smith helped us identify technology and data sources for monitoring buffers for the Water Utility Company NbS. Buffer zones are hard to identify from satellite imagery and even field margins are difficult to identify from aerial imagery. However, margins can be more obvious at different times of the year, for example, after harvest where there is bare soil for some time. However, harvest times vary a lot. Farmers have sensors and monitors in their tractors and collect data about soil moisture content. They use this to control fertilizer and pesticide rates for precision farming. This information is not public but the [National Accounts](#) report on how much fertilizer and pesticide is being used per farm every year. A little bit of grazing is good for grassland, it lets the flowers thrive, but the grass shouldn't get too low. Healthy buffers are tall and thick and this should be measurable from aerial imagery.

*Insight Investment*
We presented our model to David McNeil at Insight Investment. Insight Investment is one of the largest global asset management companies and a founding signatory of the United Nations supported Principles for Responsible Investment. David asserted the water utility company would want to lower interest rates on the bond. There would be a penalty when the water utility company fails to hit their own sustainability targets and cost of financing (rate of interest they pay) goes up. The standard 0.25% step up has no empirical basis and does not reflect the risk to the issuer. Insight would be interested in the extent to which change in the interest rate is tied to our model itself. Subsequent issuance might have different interest rates. AI could play a role in giving some step up that reflects the financial risk. Investors would be interested in this.

There is value in the predictive AI model. Risk pricing is currently based on historical judgement. Complexity of modelling is prohibitive to financial organisations. There is a case for a more empirical approach that needs to be more scientific. David was very supportive of multi-disciplinary models involving finance and ecology. Our work is addressing the same problems as them. Insight wants to buy more sustainability linked bonds but currently quality and integrity of instruments can be weak. Clients want Insight to be buying more NbS bonds.

# 4. Conclusions

Our two use cases covered contrasting examples within sustainable finance. The Brazilian Cattle Ranching Use Case, an example of greening finance, integrated nature-related considerations into mainstream financial decision-making to transition investments away from sectors with poor historical track records and unsustainable operations. The deployment of nature-based solutions in UK water utility use case, an example of financing green, drove investment to nature-positive outcomes. The two examples also covered different sectors, geographies, financial assets and AI modelling techniques, providing an overview on how AI could be applied to different challenges relating to nature's integration into finance. We provide recommendations for next steps to further enable AI solutions to accelerate nature's integration into finance for various stakeholder groups at the end of this section.

Our asset level modelling approach, with the help of AI, exploits a wide range of detailed local heterogenous data that can attribute biodiversity loss and gain to each actor in the value chain. This allows investors, and others, to see the specific impacts of investments, mitigates greenwashing and, with Generative AI[7], could provide a means for accurate, frequent, and automated disclosure. Consequently, we were able to propose E-scores for each organisation in the value chain. Uncertainty is modelled explicitly and this both informs and directs further efforts for data gathering and model refinement.

Although the approach described above represents an overall improvement in biodiversity risk measurement robustness, there are also limitations. Practically, AI techniques typically require much more computing power than more traditional methods. Furthermore, our approach requires asset-level data to achieve the necessary granularity, which has proven challenging as companies either do not have data to the required granularity or they are unwilling to disclose it voluntarily. With the Bayesian modelling approach, we would have to specifically build scenarios for scenario analysis, which might require a new model, although some scenarios can be run on the same model by tweaking initial conditions. Finally, our modelling approach is currently quite bespoke and rather involved, requiring intensive engagement and time investment on behalf of the financial institutions, supply chain partners and ecological experts, which could be challenging if not a company priority. However, by exploiting the model as a Bayesian network or graph, we can present our model in an intuitive and accessible manner with diagrams of actor relationships, which provides structure to allow meaningful engagement with experts with different areas of expertise.

The research proposal envisioned a strong component of network building across the research, business, governmental, and investment communities. It was envisioned that this network would collaboratively engage candidly in interdisciplinary conversations on existing data use, data limitations, and future needs in nature-positive investment in the UK, Brazil, and Southeast Asia. However, we experienced difficulties in engaging with the investment community, particularly the banking sector since the beginning of the project. The challenge of identifying AI-ready projects relating to biodiversity and finance should not be underestimated. Due to the interdisciplinarity of the topic, a viable use case requires people in finance, supply and ecology all interested and willing to engage at the appropriate technical level. We believe that this will happen in the near future. However, there is currently little risk incentive for banks to commit sufficient resources.

---

[7] Generative AI (or "GenAI") can produce various types of synthetic data, including text, imagery and audio from models trained on real data. ChatGPT, a natural language processing chatbot, is an example of GenAI that creates novel responses to user-input prompts.

Many financial institutions we engaged with were in the early stages of exploring their nature impacts and dependencies. The current landscape of nature discourse in financial institutions involves the definition of scope, metrics, and methodology of stocktaking and strategising the management of various nature risks. Currently, the quality of data infrastructure and level of understanding of key data analytical demands require refinement to effectively co-create specific formulae required for AI integration. Therefore, to establish these use cases, we had to delve much deeper into the ecological and financial technicalities than expected to outline their relationships to be able to identify where AI could be applied. As a result, we found ourselves generating financial models that capture the transition risks associated with investments, including reputational and financial risks. Without investor validated models, it is difficult to determine exactly how AI would be useful in each use case. With that, our models should be taken with a "pinch of salt" and, instead, represent a "flavour" of potential AI applications in this space.

Through developing the models ourselves, we were able to engage with all partners in a coherent way on the illustrated problems. With a proposed model and, particularly with the Bayesian modelling approach, we were able to engage with financial institutions and demonstrate the detail required for a successful partnership. Furthermore, we found that the model anchored the conversation and that we were able to apply each organisation's expertise in a meaningful and specific manner without requiring much cross-disciplinary debate. It was particularly challenging yet important to leverage each type of required expertise, be it ecological or financial, as we were unable to identify experts with a deep interdisciplinary understanding of nature-finance relationships.

However, to successfully implement AI solutions, we needed engagement with actors across the problem space, all contributing their portion at the required detail. In our engagements, financial institutions could not articulate what missing information was needed to integrate nature into financial decision-making. Without a clear understanding of the gaps in information required, it is not possible to identify existing data and models that might be able to fulfil their requirements and it is inappropriate to spend a lot of effort to employ potential AI solutions at this moment. The relationship between ecological and financial systems as well as the specific information requirements of financial institutions, must be better understood and articulated for AI to be an appropriate solution. As these pre-conditions have not been met our AI expert (Steven Reece) has chosen not to take a lead role in Phase 2 of the Integrating Finance and Biodiversity project. Nevertheless, he extends his availability on an *ad hoc* basis and expresses keenness to contribute whenever there are specific AI/ML needs or occasions necessitating guidance on probabilistic modelling.

*4.1 Stakeholder Recommendations*

AI can provide meaningful solutions to accelerate the integration of nature into financial decision-making.[8] However, there are some pre-conditions, as we mentioned above, that must be met to effectively employ AI successfully in the space. We have recommended next steps that relevant stakeholder groups can take to create an enabling environment for AI to be implemented and for the effective measurement and management of nature within financial systems.

---

[8] For actionable recommendations for how governments, NGOs and companies (including financial organisations) can use AI to support biodiversity conservation in general see the GPAI report *Biodiversity and Artificial Intelligence, Opportunities & Recommendations for Action* (The Global Partnership on Artificial Intelligence (GPAI), 2022).

**TNFD**
- A guidance document on AI and automation on TNFD reporting and data reporting could be helpful for financial institutions and companies, including examples of potential uses of AI in their identification of impacts and dependencies on nature. However, the development of such guidance may need to be phased.
    - There are some aspects of AI approaches that can be developed now, with the data and information currently available, such as extracting information from data and automating that process (e.g. information from satellite imagery).
    - Other aspects will be equally important but need more maturing before any guidance can be provided. This includes systems-level models that connect finance through to the value chain (e.g. attribution modelling for finance), and accurately attribute nature impact and dependency.

- Attribution modelling for finance can be used by financial institutions to understand their own risks and impacts to the ecosystem state. It could also be used by NGOs or other third-party organisations to assess the validity of public TNFD disclosures, and to ensure responsibility is taken.

**System Modellers**
- We need a *multidisciplinary* rather than interdisciplinary approach to model building. We need experts from different components that bring their expertise to the model and reach consensus with some overlap of the components in the model (e.g. connections ecological state and economic activity).

- However, we need an *interdisciplinary* view of the full value chain to make certain decisions, such as at the government-level, and for that we need a fully connected systems model (e.g. impact of legislation on the economy).

- From the AI perspective, we need a very well-defined problem description, which was something that we struggled to achieve in our use cases for this project. As the understanding of the potential use of AI grows within the industry, hopefully, the ability to articulate use cases will grow. This report gives some idea of where AI could be used in this space both in terms of data interpretation and linking system-models.

**Financial Institutions**
- Financial institutions could move forward significantly in this space by following the steps laid out in TNFD's LEAP process, and AI tools can potentially be a great help in this process:
    - Locate - FIs must determine their key interfaces with nature and where they are located. They also must identify key data sources that will provide the relevant information for the remaining LEAP components and clearly define information requirements from the data for AI solutions.

    - Evaluate - FIs must determine and define their material risks, including relevant legal frameworks, relating to nature to identify an appropriate modelling paradigm. FIs must clearly define their risk assessment and measurement process to articulate a problem definition appropriate for AI solutions.

- - Assess - As they define their processes for identifying and measuring nature-related risk, FIs must think about how AI could be employed to streamline these processes.

  - Prepare - "Generative AI" could be used to generate textual reports from complex system level models for meaningful and efficient reporting.

- There are multiple relevant timeframes for nature-related risk assessment. Nature-related risk should be considered at the "go/no-go" point of making an investment, as well as throughout the period of investment.

- FIs must be able to articulate the level of acceptable uncertainty for nature-related risk data and metrics to inform investment decisions. With a defined level of uncertainty, AI tools can then be used to identify key value chain variables that require further measurement to reduce uncertainty in the output.

**ESG Data Providers**
- To robustly include nature risks metrics within the E of ESG scores, ESG data providers need to be looking at asset-level data of the direct operations and through the value chain to understand the nature risks of a company and associated FI.

- Provide comparable and science-based methodologies to develop nature metrics that models and financial institutions can implement in their mappings and assessments.

# Appendix A: Stakeholder Contributions

Individuals from the following organisations discussed various aspects of the use cases and modelling methodology with the REDB team. The use case contributions are noted in the main body of the report. Alex Money (Smith School) helped distinguish financial return and reputation as the main investor risks for our modelling. Duncan Royle (EFTEC) commented positively on the individual E-score approach for organisations in the supply-chain, including financial organisations. He does not recommend using a probabilistic E-score even when indicator data is uncertain. Mark Fawcett (Nest Corporation) was very supportive of the approach in general and offered to connect us to RLAM. Stafan Startzel (Ernst & Young) has been contemplating a similar model approach to ours so was very supportive. Rosimiery Portela (Capitals Coalition) introduced us to [ARIES](#), a platform for interoperable data and models using Bayesian (Network) AI, that resonates with our approach. The ARIES framework focuses on quantifying the ecosystem services from the perspective of the value accrued by the beneficiaries of the services, allowing them to move from potential to actual benefits to society (Villa et al., 2014). The benefits are modelled using dynamic flow models in which services flow from "source regions" to "use regions", sometimes with sinks along the pathway absorbing or depleting service benefits (Villa et al., 2014). These regions are spatially explicitly mapped and connected in simulated Service Path Attribution Networks (SPANs) (Villa et al., 2014). The resulting flow density maps allow for the comparison between theoretical, possible and actual benefits accrued, allowing for the exploration of the efficiency of service provision and can be used for scenario analysis (Villa et al., 2014). Our approach extends the Bayesian modelling approach by explicitly including financing the ecosystem service alongside quantifying the financial benefits accrued. Interoperability is a key element of their approach and reflects the multi-disciplinary modelling paradigm of our approach. Rosimiery also mentioned climate change has affected rainfall and reduced the seasonal flooding in Brazil. Now every farm has a well to extract ground water. This is having a significant effect on the Amazon. Well pumps should be visible in high resolution imagery. Divya Narain (OxSFG) provided the background material to the Brazil Beef Supply Use case in Appendix B.

<u>Brazil Beef Farming Use case</u>
| | |
|---|---|
| TNFD | Laura Clavey, James d'Ath |
| BVRio | Grace Blackman |
| Capitals Coalition | Rosimiery Portela, Lenka Moore |
| OxSFG | Divya Narain |
| Santander Asset Management | - |
| Rabobank | - |
| JGP Asset Management | - |
| Suncorp | - |

<u>Water Utility Use case</u>
| | |
|---|---|
| Environmental Change Institute | Alison Smith |
| RLAM | Georgina Chiu |
| Insight Investment | David McNeil |

<u>AI, Modelling and E-Scores</u>
| | |
|---|---|
| EFTEC | Duncan Royle |
| Nest Corporation | Mark Fawcett |
| Ernst & Young | Stefan Wolf Startzel |
| Quant Foundry | Chris Cormack |
| Insight Investment | David McNeil |

# Appendix B: Corporate Disclosures and Controversies

Disclosures, NGO reports and newspaper articles can provide some indication of the impacts and dependencies organisations are having on nature. This information can be useful to track the organisation's reputation and also help build ESG scores. Currently, ESG scores largely rely on human analysis of company disclosures and are therefore subjective, based on backwards-looking, and often out of date data, and lack transparency in their methodology (Hughes et al., 2021). Natural Language Processing (NLP) models have been used following data scraping to generate "alternative ESG" scores, which can provide ESG scores with greater standardisation, transparency and real-time updates using a wider array of sources and conducting more objective and quantitative sentiment analysis (Hughes et al., 2021). Furthermore, NLP can be used to identify scandals or gauge reputational risks for corporates based on sentiment in the media representation. There have been greater calls for inclusion of litigation risk and, therefore, reputational risk in environmental risk assessment and management (Wetzer et al., 2024). With the immense power of NLP, sentiment can be quickly captured and quantified, therefore, including the previously illusive reputational risk with a much more systematic and robust methodology.

**Instances of Legal and reputational action against livestock companies: the case of JBS** (Acknowledgement to Divya Narain, OxSFG, for collating the following reports).

DROP JBS – A campaign website run by advocacy groups including Mighty Earth, Feedback, BankTrack and Greenpeace, urging banks and supermarkets to stop doing business with the company. Separate campaign pages for TESCO and Barclays (JBS' largest financier).

A Rotten Business: How Barclays became the go-to bank for JBS, one of the world's most destructive meat corporations – This report by advocacy organization Feedback offers a detailed chronological account of lawsuits, complaints and fines slapped against JBS for multiple offences including corruption, insider trading, price fixing, human rights violations and environmental crimes, juxtaposing it with continued support from Barclays. The report contains links to various news stories.

Another report by Institute of Agriculture and Trade Policy, 'Behind the curtain of the JBS net zero pledge' also lists various violations by JBS.

The key instances of environment-related legal and regulatory action against JBS are listed below:

- In 2017, JBS was fined 24.7 million reais (almost $8 million at a 2017 conversion rate) when IBAMA (Brazil's environmental police) discovered two JBS slaughterhouses had bought 49,468 cattle from embargoed areas.

- In 2017, the Federal Prosecutor's Office in the Amazon state of Pará published an audit finding that 19% of all JBS' cattle purchases, over 118,000 head of cattle, had failed to comply with its legally-binding no-deforestation agreements.

- In October 2021, prosecutors published audit results of JBS' compliance with its legal no-deforestation agreement, finding that over 43% of its cattle purchases in the Amazon state of Pará were not in compliance with its legal obligations.

- Numerous European supermarkets including Sainsbury's dropped products from JBS following 2021 research by Repórter Brasil that found multiple alleged examples of "cattle laundering", beef processed by JBS and sourced from "fattening" ranches that, in turn, buy from other ranches whose cattle were raised and fed on farms officially sanctioned for illegal deforestation in the Amazon or other biomes.

- In January 2023 Mighty Earth submitted a complaint to the US Securities and Exchange Commission (SEC) about a series of four 'misleading and fraudulent green bonds' that JBS issued on the back of its net-zero commitments even as the meatpacker "concealed the true scale of its emissions footprint, by failing to disclose the number of animals it slaughters every year, which are the primary source of its greenhouse gas (GHG) emissions."

- Also in 2023, after a challenge brought by the Institute for Agriculture and Trade Policy, the National Advertising Review Board (the US's advertising watchdog) asked JBS to stop making "aspirational" net-zero claims that are not backed by actual measures.

The reports above pertain to *corporate controversies* and contribute to the reputation risk of the investor, as well as organisations in the value chain. We indicate how these reports can be folded into our probabilistic reputation model in Section 2.4.4 Reputation Loss.

Formally, we require the probability that an organisation complies with the Forest Code, or has a particular NRP impact score, given both the situation data $D$ from Section 3 and reports $R$. The addition of $R$ should provide a clearer picture of the reputational risk for the investor than $D$ alone.

Without going into too much detail, we can extend our Bayesian model to accommodate the reports and provide some detail of how the reputational risk is calculated. We will focus on the Forest Code compliance, $L$, but note our analysis applies to the NRP impact score, by replacing $L$ with $N$ in the equations.

We assign a class $s$ to each report and then use this value in our Bayes model to update the reputational risk of the investor. The class could indicate the *sentiment* of nature-related developments in the report, for example, *negative risk*, *neutral* or *positive opportunity,* or could be more explicit, *controversy reported, not reported* etc. Initially humans will classify some reports by hand to train a large language classifier $p(s \mid R)$. The model $p(s \mid R)$ could comprise large language models BERT or GPT-2 with an appropriate downstream classifier (Sun et al., 2023; Webersinke et al., 2022). This classifier is then used to label all reports. The most probable class, $s_{max,R} = \text{argmax}_s\, p(s \mid R)$ is assigned to report $R$.

We assume the document sentiment class is correlated with the reported (or reporting) organisation's compliance with the Forest Code. This is captured via the factor $p(s_{max,R} \mid L)$ which can be integrated into our reputation model using Bayes rule to provide the posterior belief in the Forest Code compliance given both $D$ and $R$:

$$p(L \mid B, D, R) \propto p(s_{max,R} \mid L)\, p(L \mid B, D).$$

# Appendix C: Acknowledgements

We acknowledge, and are grateful for, the support of the following who kindly proofread this report and offered some very insightful suggestions.

| | |
|---|---|
| Calvin Quek | Smith School, Oxford University |
| Christophe Christiaen | Smith School, Oxford University |
| Alok Singh | Smith School, Oxford University |
| Divya Narain | Smith School, Oxford University |
| Olga Isupova | Leverhulme Centre for Nature Recovery, Oxford University |

**Oxford Sustainable Finance Group**
Smith School of Enterprise and the Environment
& **Leverhulme Centre for Nature Recovery**
School of Geography and the Environment
University of Oxford
South Parks Road
Oxford, OX1 3QY
United Kingdom

E: steven.reece@ouce.ox.ac.uk
E: emma.odonnell@biology.ox.ac.uk

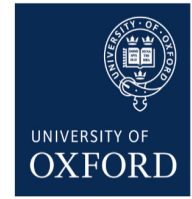

**Department of Environment & Geography**
University of York
Wentworth Way, Heslington
York, YO10 5NG
United Kingdom

E: felicia.liu@york.ac.uk

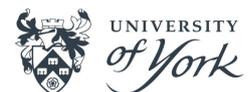

**UNEP-WCMC**
219 Huntingdon Road
Cambridge
CB3 0DL
United Kingdom

E: joanna.wolstenholme@unep-wcmc.org
E: frida.arriaga@unep-wcmc.org
E: giacomo.ascenzi@unep-wmcm.org

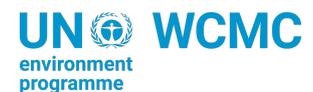

**UK Centre for Ecology and Hydrology**
Maclean Building
Benson Lane
Crowmarsh Gifford
Wallingford
OX19 8BB
United Kingdom

E: rfp@ceh.ac.uk (Richard Pywell)

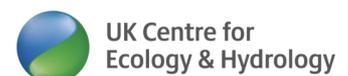